\documentclass[review,5p]{elsarticle}
\journal{Physica D}
\usepackage{hyperref}
\usepackage[utf8]{inputenc}
\usepackage{amsmath}
\usepackage{amssymb}

\usepackage{graphicx}
\usepackage[american]{babel}
\usepackage[amssymb]{SIunits}
\usepackage[caption=false]{subfig}
\usepackage{notoccite}
\usepackage{afterpage}
\newcommand{\vphi}{\mbox{\boldmath{$\phi$}}}

\newcommand{\vmu}{\mbox{\boldmath{$\mu$}}}
\usepackage{tikz}
\usepackage{ulem}
\usetikzlibrary{calc}
\usetikzlibrary{plotmarks}
\usepackage{pgfplots}
\usepackage{etoolbox}
\patchcmd{\normalsize}{13.6}{13}{}{}
\pgfplotsset{
tick label style={font=\tiny},
label style={font=\small},
legend style={font=\tiny}
}

\begin{document}

\begin{frontmatter}

\title{Revisiting Jackson-Hunt calculations: Unified theoretical analysis for generic multi-phase growth in a multi-component system}

\author[add]{Arka Lahiri\corref{correspondingauthor}}
\ead{arka@platinum.materials.iisc.ernet.in}

\author[add]{Abhik Choudhury}%
\ead{abhiknc@materials.iisc.ernet.in}

\address[add]{%
Department of Materials Engineering, 
Indian Institute of Science, Bangalore - 560 012, India.
}%
\cortext[correspondingauthor]{Corresponding author}

\begin{keyword}
Phase-field; Jackson-Hunt; multi-component; multi-phase; multi-variant; in-variant; eutectic
\end{keyword}

\begin{abstract}
A straight-forward extension of the Jackson-Hunt theory for directionally solidifying multi-phase growth
where the number of components exceeds the number of solid phases becomes difficult on account
of the absence of the required number of equations to determine the boundary layer compositions ahead
of the interface. In this paper, we therefore revisit the Jackson-Hunt(JH) type calculations for any 
given situation of multi-phase growth in a multi-component system and self-consistently derive the variations
of the compositions of the solid phases as well as their volume fractions, which grow
such that the composite solid-liquid interface is isothermal. This allows us to unify the (JH) calculation
schemes for both in-variant as well as multi-variant eutectic reactions. The derived analytical expressions 
are then utilized to study the effect of dissimilar solute diffusivities and interfacial energies
on the undercoolings and the solidified fractions. We also perform phase field simulations to confirm 
our theoretical predictions and find a good agreement between our analytical calculations and model predictions 
for model symmetric alloys as well as for a particular Ni-Al-Zr alloy. 
\end{abstract}

\end{frontmatter}


\section{Introduction}
Eutectic solidification in a generic multi-component alloy, where two or more solids exhibit coupled growth, 
can be associated with degrees of freedom greater than or equal to zero. Experimentally, invariant (zero degrees of freedom)  
eutectic reactions 
have been observed in~\cite{Kerr1964,Cooksey1967,Bao1972,Rinaldi1972,Holder1974,Ruggiero1995,Hecht2004,
Rex2005,Witusiewicz2006,Contieri2008,Genau2012,Dennstedt2012,Dennstedt2013}
Our theoretical understanding of invariant eutectic reactions 
is fairly advanced for binary~\cite{Brandt1945,Zener1946,Hillert1957,Jackson1966} 
as well as for ternary systems~\cite{Himemiya1999,Choudhury2011}. 
In these studies, the solid phase fractions are assumed to be the ones predicted by the equilibrium phase  
diagram at the invariant temperature. This allows the determination of the magnitude of the composition 
boundary layers at the eutectic front from the conditions of equality of undercoolings at different
solid-liquid interfaces.

In contrast to binary alloys, multi-component alloy systems can display eutectic 
reactions which are not invariant (possessing degrees of freedom greater than zero).
An example of such a reaction is the concurrent solidification of two phases 
in ternary alloys, which possesses a single degree of freedom, and is known as the
monovariant (or univariant) eutectic. This reaction exists
over a range of temperatures in the equilibrium phase diagram 
compared to its invariant counterpart in binary. 
Experimental studies in several multi-component systems report the existence of such multivariant
reactions~\cite{DeWilde2004,Yamauchi1996,Rinaldi1972,Garmong1971,Fehrenbach1972,Schnake1997,Raj2001}.
Multivariant eutectics are also susceptible to Mullins-Sekerka~\cite{Mullins1964} like destabilization of the 
solidification front in the presence of one or more impurity components leading to the
formation of eutectic cells or colonies as seen in~\cite{Akamatsu2000,Akamatsu2010}. Eutectic cells 
have been studied theoretically and numerically in~\cite{Plapp1999} and~\cite{Plapp2002}, respectively.

The theoretical development for multi-variant eutectic reactions have mainly
been centered around ternary monovariant eutectics. 
An attempt to extend the JH-type calculations to explain monovariant eutectics leads
to an under-determined system where 
the number of unknowns (for describing the magnitude 
of the solute boundary layers) exceed the equations (the equality of interfacial 
undercoolings at different solid-liquid interfaces).
This marks a departure from the theories of invariant eutectics~\cite{Jackson1966,Himemiya1999,Choudhury2011},
where there are enough equations as unknowns to render the system consistent.

McCartney et al.~\cite{Mccartney1980} are the first to circumvent  
this difficulty by introducing an additional constraint relating the magnitude of the composition
boundary layers of the two independent components assuming the solid phase fractions to be given by the equilibrium 
phase diagram.
The under-determined nature of the problem of ternary monovariant eutectics urge 
a reduction in the number of the unknowns in the problem by expressing them as functions 
of the solid phase volume fractions. Thus, the solid phase fractions are no longer 
determined by the equilibrium phase diagram but by the growth dynamics.
The dynamic selection of solid phase fractions during growth being an experimental fact
valid for invariant and non-invariant eutectics alike, prompted several theoretical studies
which attempt to understand eutectic solidification dynamics as functions of solid phase 
fractions. Donaghey and Tiller~\cite{Donaghey1968} calculate the composition fields in the liquid 
as functions of the solid phase fractions for binary as well as ternary systems, which are
then utilized by Ludwig and Leibbrandt~\cite{Ludwig2004} to demonstrate the dependence
of interfacial undercoolings on  solid phase fractions, but for binary systems only.
Magnin and Trivedi~\cite{Magnin1991} consider different densities of the 
eutectic solid phases for a binary eutectic to derive the curvature 
at each point of the eutectic solidification front required to maintain an isothermal solid-liquid interface.
These curvatures when averaged over the two eutectic solids, lead to contact angles 
different from what is predicted by the criterion of mechanical equilibrium at the triple points,
constituting a driving force for a dynamic selection of volume fractions of 
solid phases different from what is predicted by
the equilibrium phase diagram. For ternary monovariant eutectics, 
De Wilde et al.~\cite{DeWilde2005} treat a 
particular solid phase fraction as a material parameter 
which is then varied to obtain the corresponding variation 
in the  growth dynamics at the extremum condition
of minimum undercooling of the eutectic front.

A couple of recent theoretical studies have a gone a step further 
by presenting a method to compute the solid phase fractions as
a key step to determine non-invariant eutectic growth involving 
two solid phases in a generic multi-component alloy. 
The study by Catalina et al.~\cite{Catalina2015} 
presents a linearized theory in this regard without allowing  
for the changes in composition in one of the solidifying phases. 
Here, the solid phase fractions are determined from the criterion 
of equal undercoolings at the two solid-liquid interfaces.
A more rigorous extension is provided recently 
by Senninger and Voorhees~\cite{Senninger2016},
where they take into account the composition variations of both the solid phases.
Although, our work shares a similar spirit in this aspect, 
we present an alternate derivation. One of the major differences
is that we relate the deviations of the phase compositions to the 
departures of the diffusion potentials and temperature and thereby
the functional dependence between the variations of the solid 
and liquid compositions is more elegantly retrieved.
Secondly, our theory is applicable for any generic
multi-phase eutectic growing with a lamellar arrangement which 
is in contrast to the work of  Senninger and Voorhees~\cite{Senninger2016}, 
who limit themselves to two-phase growth.
In addition, we verify our analytical calculations
with phase-field simulations considering  model symmetric alloys as well as 
a Ni-Al-Zr alloy. In all the studies mentioned above, 
the effect of solute diffusivities in modifying the 
selection of solid phase fractions have not been explored.
We explore this aspect using our phase-field simulations
as well as analytical calculations.

\section{Analytical theory}

\subsection{The Jackson Hunt calculation}
In order to motivate our present work let us 
re-visit the main results of the classical Jackson-Hunt
analysis as detailed in \cite{Jackson1966}, for deriving the 
undercooling vs spacing relationships for two-phase growth in a binary
alloy. The situation is modeled by considering a repeating representative 
unit of two phases $\alpha$ and $\beta$ growing in a directional 
solidification set-up where the imposed temperature gradient (G) at the 
interface traverses with a velocity V, that sets the rate of solidification.
The undercooling at each interface can be written as, 

\begin{align}
\Delta T^{\nu} = 
-m_B^\nu(\widetilde{c_B^\nu} - c_B^E) + \Gamma_\nu \widetilde{\kappa_\nu},
\label{clos_cond_bin}
\end{align}
where, $\widetilde{c_B^\nu}$ represents the average composition in the liquid in local equilibrium with 
the $\nu$-th phase and $c_B^E$ represents the eutectic composition. $m_B^\nu$ is the liquidus slope. 
$\Gamma_\nu$ and $\kappa_\nu$ denote the Gibbs-Thomson coefficient and the interfacial curvature, respectively.

We start by writing the composition profiles as a Fourier 
series with amplitudes that are determined from the condition 
that the composition profiles obey both the governing equation 
and the Stefan condition. A corresponding generic analysis for invariant eutectic growth
in  multi-component systems is laid out in~\cite{Choudhury2011}, where expressions 
for all the amplitudes apart from the zeroth order mode 
(representing the boundary layer) can be determined from the 
inverse Fourier transform. In order to understand the difficulty in determining the amplitude of the zeroth order, 
we first inspect the expression obtained by performing an inverse Fourier transform of the same,
written as, $B_0 = \left[c_B^{\alpha l}\eta_\alpha + c_B^{\beta l} (1-\eta_\alpha)\right] 
- \left[c_B^\alpha \eta_\alpha + c_B^\beta (1-\eta_\alpha)\right]$,
where $c_B^{\alpha/\beta, l}$ represents the liquid compositions
in equilibrium with the $\alpha/\beta$ interfaces, $\eta_\alpha$
is the volume fraction of the $\alpha$ phase and $B_0$ denotes the boundary 
layer amplitude corresponding to the component $B$.
If one uses the volume fractions and compositions at the eutectic
temperature (as for the other Fourier modes) 
for determining the boundary layer composition $B_0$ (far-field 
composition is at the eutectic) it would
result in zero and the corresponding undercoolings at the interface would not be
equal. This calculation, would also be physically incorrect, 
as the phase compositions deviate from their values at
the eutectic temperature. Jackson and Hunt in their analysis, 
treat this difficulty by keeping the $B_0$ as an unknown 
which is fixed by the condition that the undercoolings
at both solid-liquid interfaces are equal, while the volume 
fractions $\eta_{\alpha}$ at the eutectic temperature are 
utilized for computing both the constitutional and curvature 
undercoolings. In general, one can solve this problem of 
invariant growth for a multi-component system as in \cite{Choudhury2011},
where it has been shown to agree well with experiments
as well as phase-field simulations.

For the mono-variant reaction however, for instance in a two-phase growth in a ternary alloy, 
there would be two boundary layer compositions, whereas the 
equality of a common undercooling imparts only a single equation,
thereby the system of equations become under-determined. The system 
of equations can only be made deterministic by invoking the functional 
dependence of the boundary layer compositions on the phase
compositions and the solid-fractions. This motivates our present derivation, 
which in spirit unifies the theories of in-variant and multi-variant eutectic growth.

\subsection{Theory}
The following discussion is generic to a directionally solidifying 
multi-component alloy of $K$ components (with $K-1$ of them being independent), 
displaying a eutectic reaction with $N$ solid phases, possessing a degree of freedom given by $F=K-N$. 
Though, we present the theory assuming 
independent diffusion of solutes in the liquid (no diffusion in the solid), 
it can be considered to be representative of a system with non-zero off-diagonal 
terms in the diffusivity matrix, when such an analysis is carried out in the basis system of the eigenvectors
of the diffusivity matrix. In the following discussion, 
the indices $i$ and $j$ are reserved for
solutes, while $\nu$ and $p$ denote the solid phases appearing due to eutectic solidification.  
Assuming a flat interface, the composition variation in the liquid is of the form~\cite{Choudhury2011},
 \begin{align}
  c_i=c_i^\infty+\sum_{n=-\infty}^{n=\infty}I_n e^{\hat{i} k_n x-q_n^i z},
  \label{comp_var}
 \end{align}
 where, $\hat{i}=\sqrt{-1}$, and $k_n=2\pi n/\lambda$, 
 are wavenumbers characterizing the variation of solute concentrations in the liquid across a 
 solid-liquid interface aligned along $x$ with the eutectic solids growing in $z$.
 Conformity of Eq.~\ref{comp_var} to the stationary form of the diffusion equation given below,
 \begin{align}
  V\dfrac{\partial c_i}{\partial z} + D_{ii} \nabla^2 c_i=0,
 \end{align}
leads to,
 \begin{align}
  q_n^i=\left(\dfrac{V}{2 D_{ii}}\right) + \sqrt{k_n^2 + \left(\dfrac{V}{2 D_{ii}}\right)^2},
  \label{qn_def}
 \end{align}
 where $D_{ii}$ denotes the diffusivity of the $i$th component and $V$ represents the sample pulling velocity.  
 Following the discussion in~\cite{Choudhury2011},
 a single wavelength of the eutectic consists of $M$ units ($M>=N$) of the 
 eutectic solids with each one of the $M$ units corresponding to one of the $N$ phases.   
 The periodic variation starts at $x_0=0$ and terminates at $x_{M}=1$ with the 
 width of the $\nu$-th unit being given by $\left(x_{\nu}-x_{\nu-1}\right)\lambda$;
 the entire wavelength being $\left(x_M-x_0\right)\lambda$. 
 Thus, the volume fraction of a particular phase $p$, denoted by $\eta_p$, 
 can be calculated from a single wavelength of the eutectic lamellae as,
 \begin{align}
  \eta_p=\sum_{\nu=0}^{M-1} \left(x_{\nu+1} - x_{\nu}\right) \delta_{\nu p},
  \label{eta_def}
 \end{align}
 where,
 \begin{align}
  \delta_{\nu p} =
    \begin{cases}
            1, &         \text{if } \nu=p,\\
            0, &         \text{if } \nu\neq p.
    \end{cases}
    \label{kron_delta}
 \end{align}

 The mass balance across a particular location at the solid-liquid interface for the $\nu$-th unit can be written as,
 \begin{align}
  V\Delta c_i^{\nu}=-D_{ii}\dfrac{\partial c_i}{\partial z}\Bigg{|}_{z=0,{x_\nu^*}},
  \label{stefan_cond}
 \end{align}
 with $\Delta c_i^{\nu}= c_i^{\nu l} - c_i^{\nu}$, where $c_i^{\nu l}$ and $c_i^{\nu}$ denote the 
 liquid and the solid compositions in local equilibrium at a location $x_\nu^*$ on the $\nu-l$ interface, respectively.
 An expression for the Fourier constants $I_n$ is obtained by 
 invoking the orthogonality of the Fourier basis functions while 
 integrating Eq.~\ref{stefan_cond} over an entire period ($\lambda$) 
 of the eutectic, leading to,
 \begin{align}
  q_n^i I_n \lambda = \dfrac{2}{l_i}\sum_{\nu=0}^{M-1} \int_{x_\nu\lambda}^{x_{\nu+1}\lambda} \exp(-ik_n x) \Delta c_i^{\nu} dx,
  \label{fou_const}
 \end{align} 
 where, $l_i=2 D_{ii}/V$, is the diffusion length associated with the $i$-th component.
 For the mode corresponding to $n=0$, Eq.~\ref{fou_const} yields for the $i$-th component,
 \begin{align}
 I_0 = \dfrac{1}{\lambda}\sum_{\nu=0}^{M-1} \int_{x_\nu\lambda}^{x_{\nu+1}\lambda} \Delta c_i^{\nu} dx.
 \label{def_I0_org}
 \end{align}
 It is beneficial to define average compositions in front of a particular 
 phase $p$ as, 
 \begin{align} 
 \overline{\Delta c_i^{p}}=\dfrac{\sum_{\nu=0}^{M-1} \delta_{\nu p} \int_{x_\nu \lambda}^{x_{\nu+1}\lambda}  \Delta c_i^{\nu} dx}  
 {\sum_{\nu=0}^{M-1}\left(x_{\nu+1}-x_\nu\right)\lambda \delta_{\nu p} }, 
 \label{wt_avg_del_c}
 \end{align}
 which allows us to re-express Eq.~\ref{def_I0_org} as,
  \begin{align}
 I_0 = \sum_{p=1}^{N} \eta_p \overline{\Delta c_i^{p}}.
 \label{def_I0}
 \end{align}
Similarly, the average interfacial composition in the solid ($\overline{c_i^{p}}$) and the liquid $\overline{c_i^{pl}}$ ahead of it is defined as,
\begin{align} 
 \overline{c_i^{p,pl}}=\dfrac{\sum_{\nu=0}^{M-1} \delta_{\nu p} \int_{x_\nu \lambda}^{x_{\nu+1}\lambda}  c_i^{\nu,\nu l} dx}  
 {\sum_{\nu=0}^{M-1}\left(x_{\nu+1}-x_\nu\right)\lambda \delta_{\nu p} }.
 \label{wt_avg_sol_liq_c}
\end{align}
The theory in~\cite{Choudhury2011} following that of Jackson \& Hunt~\cite{Jackson1966} 
provides an expression for $\overline{c_i^{pl}}$, which has the form, 
\begin{align}
 \overline{c_i^{pl}}&=c_i^\infty+I_0 \nonumber \\
 &+ \dfrac{\lambda}{\eta_p l_i} f_i\Big(P_1(\eta_1,\cdots,\eta_N),\cdots,P_r(\eta_1,\cdots,\eta_N), \nonumber \\
 &\overline{\Delta c_i^{1}},\cdots,\overline{\Delta c_i^{N}} \Big),
 \label{JH_avg_comp}
\end{align}
where, each of one of the $k$ infinite series' $P_k(\eta_p)$, $k=1,\cdots,r$, $p=1,\cdots,N$, 
are composed of terms which are trigonometric functions of $\eta_p$. The value of $r$ and the form of $P_k(\eta_p)$
are determined by the number and repetitions of solid phases in a single periodic unit of wavelength $\lambda$. It must be 
mentioned at this point that the term $I_0$ represents the principal term determining the liquid compositions $\overline{c_i^{pl}}$ 
at the flat interface,
with the secondary influence being due to that of the higher order modes averaged over the lamellar widths denoted by the final term in the 
RHS of Eq.~\ref{JH_avg_comp}. An example of such a term for a ternary monovariant eutectic~\cite{Choudhury2011},
\begin{align}
 f_i&=2 P(\eta_\alpha) \left(\overline{\Delta c_i^\alpha} - \overline{\Delta c_i^\beta} \right),  \\
 P(\eta_\beta)&=P(\eta_\alpha)= \sum_{n=1}^\infty \dfrac{1}{(\pi n)^3} \sin^2(\pi n \eta_\alpha).
 \label{f_tme}
\end{align}
The average undercooling ($\overline{\Delta T^p}$) ahead of a particular solid ($p$)-liquid ($l$) interface is given by,
\begin{align}
 \overline{\Delta T^p}= T^* - T^p = \sum_{i=0}^{K-1} m_i^{p}\left(c_i^{l,*} - \overline{c_i^{pl}} \right) + \Gamma_p\overline{\kappa_p},
 \label{delT_ahead_of_p}
\end{align}
where, $m_i^{p}$ are the liquidus slopes, $\Gamma_p$ denotes the Gibbs-Thomson coefficient with the average
curvature of the solid($p$)-liquid interface ($\overline{\kappa_p}$)
given by,
\begin{align}
 \overline{\kappa_p} &= \dfrac{2 \sin \overline{\theta_{pm}}}{\eta_p\lambda},
 \label{kappa_def}
\end{align}
where, $\overline{\theta_{pm}}$ is the angle made by the tangent to the solid($p$)-liquid($l$) interface and 
the horizontal towards the side of the $p$-th phase when located adjacent to the $m$-th phase, and averaged over 
all such contiguous arrangements of the solid phases $m$ and $p$ in the entire period.

The fact that the imposed thermal gradient has a length scale much larger than the lamellar width, implies the
growth of all the eutectic solids at equal undercoolings, which can be expressed as,
\begin{align}
 \overline{\Delta T^1} = \overline{\Delta T^2} =\cdots=\overline{\Delta T^N} =\overline{\Delta T}.
 \label{equal_undercool}
\end{align}
Also, the sum of volume fractions of the phases in a single period of the lamellae must be equal to unity,
\begin{align}
 \sum_{p=1}^N \eta_p=1.
 \label{sum_eta}
\end{align}

Here, one needs to solve 
the Eqs.~\ref{def_I0},~\ref{JH_avg_comp},~\ref{delT_ahead_of_p},~\ref{equal_undercool} and 
~\ref{sum_eta} simultaneously, in order to retrieve the phase
compositions, the solid fractions and the undercooling of the eutectic
growth front. For this, we need to describe the functional dependence
between the solid compositions $\overline{c_i^{p}}$ and the liquid
compositions $\overline{c_i^{pl}}$. We do this by calling upon the 
relations of the phase compositions $\overline{c_i^{p,pl}}\left(\overline{\mu_{i}^{p}},T\right)$, 
where $\overline{\mu_{i}^{p}}$ corresponds to the diffusion potential of $p-l$ equilibrium,
averaged over all occurrences of the solid phase $p$ in a periodic unit of the eutectic. This then reduces the above system, 
Eqs.~\ref{def_I0},~\ref{JH_avg_comp},~\ref{delT_ahead_of_p},~\ref{equal_undercool} and~\ref{sum_eta}
in terms of $\overline{\mu_{i}^{p}}$, $T$, and $\eta_p$.

A point to note here is that, 
Senninger and Voorhees \cite{Senninger2016},
replace the Eqs.\ref{def_I0} with a mass conservation constraint.
Mass conservation is implicit in our set of equations. This can be 
seen by considering only those $N$ equations out of the $N(K-1)$ in Eqs.\ref{JH_avg_comp} 
which represent the composition fields of a particular component $i$
in the liquid in equilibrium with different solids ($p$). Summing over all such equations 
after multiplying both sides of each of them 
with the respective volume fractions $\eta_p$, gives
$I_0 = \sum_{p}\overline{c_i^{pl}}\eta_p - c_i^{\infty}$.
This along with Eqs.\ref{def_I0} implies that 
$\sum_{p}\overline{c_i^{p}}\eta_p=c_i^{\infty}$, which is the 
mass conservation equation used by Senninger and Voorhees~\cite{Senninger2016}. 

However, given that the thermodynamical relations \\ 
$\overline{c_i^{p,pl}}\left(\overline{\mu_{i}^{p}},T\right)$ are routinely 
non-linear, the resultant set of equations become difficult
to resolve. Given that most departures from equilibrium
in case of eutectic reactions are small, we therefore 
linearize our set of equations about the chosen eutectic temperature $T^{*}$, 
the average diffusion potentials $\mu_j^{p,*}$ and 
the solid phase fractions $\eta_p^*$, corresponding to $T^*$. 
It must be noted that the equilibrium solid and liquid phase compositions
corresponding to $T^*$, $\mu_j^{p,*}$ and $\eta_p^*$,
are $c_i^{p,*}$ and $c_i^{pl,*}$, respectively.
This results in a set of linear equations which can be 
solved for, consistently.

\subsection{Linearized Theory}
We express the average compositions in the solid ($\overline{c_i^p}$) 
and the liquid ($\overline{c_i^{pl}}$) as functions of diffusion potentials ($\overline{\mu_j^{p}}$) as,
\begin{align}
 \overline{c_i^p}=c_i^{p,*} + \sum_{j=1}^{K-1}\left[\dfrac{\partial c_i^p}{\partial \mu_j}\right]_{\mu_j^{p,*}} \overline{\Delta \mu_j^{p}}
 - \dfrac{\partial c_i^p}{\partial T}\Bigg{|}_{T^*} \overline{\Delta T^{p}},
 \label{gen_avg_sol_comp}
\end{align}
and,
\begin{align}
 \overline{c_i^{pl}}=c_i^{pl,*} + \sum_{j=1}^{K-1}\left[\dfrac{\partial c_i^l}{\partial \mu_j}\right]_{\mu_j^{p,*}}\overline{\Delta \mu_j^{p}}
 - \dfrac{\partial c_i^{l}}{\partial T}\Bigg{|}_{T^*} \overline{\Delta T^p},
 \label{gen_avg_liq_comp}
\end{align}
where, $\overline{\Delta \mu_j^{p }}= \overline{\mu_j^{p}} - \mu_j^{p,*}$, denote the change in 
average diffusion potentials from their values at the chosen eutectic point $\mu_j^{p,*}$,
under an additional constraint of constant $\left[\partial c_i/\partial \mu_j\right]$ matrices. 
The vector $\partial c_i^{p}/\partial T$ determine the change in composition with temperature at constant diffusion potentials.

Employing Einstein's indicial notation which conveys summation over repeated indices (except for $p$ in our analysis,
which denotes a particular phase),
the above equations can be written as,
\begin{align}
 \overline{c_i^{p}}&=  c_i^{p,*}  + \chi_{ij}^p 
\overline{\Delta \mu_j^{p}} - \zeta_i^p \overline{\Delta T^{p}}, 
\label{gen_avg_sol_comp_mat_mot}
\end{align}
and, 
\begin{align}
  \overline{c_i^{pl}}&=  c_i^{pl,*}  + \chi_{ij}^l 
\overline{\Delta \mu_j^{p}} - \zeta_i^l \overline{\Delta T^{p}},
\label{gen_avg_liq_comp_mat_not}
\end{align}
where,
\begin{align}
\chi^{p}_{ij} &= \dfrac{\partial c_i^p}{\partial \mu_j}\Bigg{|}_{\mu_j^{p,*}},  \\
\chi^{l}_{ij} &= \dfrac{\partial c_i^l}{\partial \mu_j}\Bigg{|}_{\mu_j^{p,*}},  \\
\zeta^{p}_i &= \dfrac{\partial c_i^p}{\partial T}\Bigg{|}_{T^*}, \\ 
\zeta^{l}_i &= \dfrac{\partial c_i^l}{\partial T}\Bigg{|}_{T^*}.
\label{not1}
\end{align}

Thus, the difference in the average compositions of the solid and the liquid as obtained from 
Eqs.~\ref{gen_avg_sol_comp},~\ref{gen_avg_liq_comp},~\ref{gen_avg_sol_comp_mat_mot} and~\ref{gen_avg_liq_comp_mat_not},
\begin{align}
 \overline{\Delta c_i^{p}} &= \overline{c_i^{pl}} - \overline{c_i^{p}}, \nonumber \\
 &= \left(c_i^{pl,*} - c_i^{p,*}\right) + 
 \sum_{j=1}^{K-1}\left(\dfrac{\partial c_i^l}{\partial \mu_j} - \dfrac{\partial c_i^p}{\partial \mu_j}\right)_{\mu_j^{p,*}} \overline{\Delta \mu_j^{p }}\nonumber \\
 &- \left( \dfrac{\partial c_i^{l}}{\partial T} - \dfrac{\partial c_i^{p}}{\partial T} \right)_{T^*} \overline{\Delta T^p} \nonumber \\
 &= \Delta c_i^{p,*} + \Delta\chi^{p}_{ij}
 \overline{\Delta \mu_j^{p}} - \Delta\zeta^{p}_i \overline{\Delta T^{p}},
 \label{gen_diff_avg_liq_sol}
\end{align}
where, to obtain the last equality expressed in indicial notation, we have used,
\begin{align}
 \Delta c_i^{p,*} &= c_i^{pl,*} - c_i^{p,*},   \\
 \Delta \chi_{ij}^{p} &= \chi_{ij}^l-\chi_{ij}^p, \\
 \Delta\zeta^{p}_i &= \zeta^l_i - \zeta^{p}_i.
 \label{diff_prop}
 \end{align}
Using, Eqs.~\ref{gen_avg_sol_comp_mat_mot},~\ref{gen_avg_liq_comp_mat_not} and~\ref{gen_diff_avg_liq_sol}, the $2N(K-1)$ composition variables 
have been expressed as functions of $N(K-1)$ intensive variables in the form of change in diffusion potentials $\overline{\Delta \mu_j^{p }}$,
which can be further related to $\Delta \eta_p (=\eta_p - \eta_p^*)$
and $\overline{\Delta T^p}$ by invoking equality of Eqs.~\ref{JH_avg_comp} and ~\ref{gen_avg_liq_comp} which provide
additional $N(K-1)$ equations, stated in the indicial notation (with no sum over $p$ and $i$) as,
\begin{align}
 \overline{c_i^{pl}}&=  c_i^{pl,*}  + \chi_{ij}^l 
\overline{\Delta \mu_j^{p}} - \zeta_i^l \overline{\Delta T^{p}}\nonumber \\
&=c_i^\infty+I_0 \nonumber \\
&+ \dfrac{\lambda}{\eta_p l_i} f_i\Big{(}P_1(\eta_1,\cdots,\eta_N), \cdots,P_r(\eta_1,\cdots,\eta_N), \nonumber \\
&\overline{\Delta c_i^{1}}, \cdots, \overline{\Delta c_i^{N}}\Big{)}
\label{liq_comp_eq}
 \end{align}
The RHS of Eq.~\ref{liq_comp_eq} (or Eq.~\ref{JH_avg_comp}), is in general 
non-linear in $\overline{\Delta \mu_j^{p}}$, $\Delta \eta_p$ and $\overline{\Delta T^p}$.
Thus, to express $\overline{\Delta \mu_j^{p}}$ as an explicit function
of $\Delta \eta_p$ and $\overline{\Delta T^p}$, we linearly
expand each term in the RHS of Eq.~\ref{liq_comp_eq} starting with $I_0$, given by, 
\begin{align}
I_0&= I_0^*+\sum_{m=1}^N\dfrac{\partial I_0}{\partial \overline{\Delta c_i^{m}}}\sum_{j=1}^{K-1}
\dfrac{\partial \overline{\Delta c_i^{m}}}{\partial \mu_j}\Bigg{|}_{ \mu_j^{m, *}} \overline{\Delta \mu_j^{m }} \nonumber \\  
&-\sum_{m=1}^N\dfrac{\partial I_0}{\partial \overline{\Delta c_i^{m}}}\dfrac{{\partial \overline{\Delta c_i^{m}}}}{\partial T}\Bigg{|}_{T^*}
\overline{\Delta T^m}\nonumber \\
&+\sum_{m=1}^{N}\dfrac{\partial I_0}{\partial \eta_m}
\Bigg{|}_{\eta_m^*} \Delta \eta_m,
\label{Taylor_I0}
\end{align}
where, $\mu_i^{p, *}$, $T^*$ and $\eta_p^*$ are the quantities corresponding to the equilibrium in the phase diagram. 
The different terms in the RHS of Eq.~\ref{Taylor_I0} can be computed from Eqs.~\ref{def_I0},~\ref{gen_diff_avg_liq_sol} and~\ref{diff_prop} as,
 \begin{align}
  \dfrac{\partial I_0}{\partial \overline{\Delta c_i^{p}}} &=  \eta_p^*, \\
  \dfrac{\partial \overline{\Delta c_i^{p}}}{\partial \mu_j}\Bigg{|}_{\mu_j^{p,*}}&= \left(\dfrac{\partial c_i^l}{\partial \mu_j} - 
  \dfrac{\partial c_i^p}{\partial \mu_j}\right)_{\mu_j^{p,*}} = \Delta \chi_{ij}^p, \\
  \dfrac{\partial \overline{\Delta c_i^{p}}}{\partial T}\Bigg{|}_{T^*} &= \left( \dfrac{\partial c_i^{l}}{\partial T} - 
  \dfrac{\partial c_i^{p}}{\partial T} \right)_{T^*} = \Delta \zeta_i^p,\\
  \dfrac{\partial I_0}{\partial \eta_p}\Bigg{|}_{\eta_p^*} &= \Delta c_i^{p, *}.
 \label{dI0_dmu}
 \end{align}
 At this point we introduce the following quantities (no sum over $p$),
 \begin{align}
  \overline{\Delta \chi_{ij}^p} &=\eta_p^* \Delta \chi^p_{ij}, \nonumber \\
  \overline{\Delta \zeta_i^p} &=\eta_p^*\Delta \zeta^p_{i},
  \label{diff_frac_vol_avg}
 \end{align}
to express Eq.~\ref{Taylor_I0} in indicial notation as,
\begin{align}
 I_0= I_0^*+ 
 \overline{\Delta \chi_{ij}^m} \overline{\Delta \mu_j^{m}} - \overline{\Delta \zeta_i^m} \overline{\Delta T^m}
 + \Delta c_i^{m,*} \Delta \eta_m,
 \label{Taylor_I0_cond}
\end{align}
with $m$ and $j$ being the indices representing phases and components respectively, which are summed over. 
The second term in the RHS of Eq.~\ref{liq_comp_eq}, being only a second-order correction to the interfacial liquid composition
is assumed to be a function of $1/\eta_p$ only, with all the other quantities evaluated at the conditions prevailing at the eutectic.
This simplifying assumption is necessary to maintain tractability of the equations. 
Thus, writing,
\begin{align}
 f_i&= f_i^*\Big{(}P_1(\eta_1^*,\cdots,\eta_N^*), \cdots,P_r(\eta_1^*,\cdots,\eta_N^*), \nonumber \\
 &\overline{\Delta c_i^{1,*}}, \cdots, \overline{\Delta c_i^{N,*}}\Big{)},
 \label{f_const}
\end{align}
we re-write the linearized version of Eq.~\ref{liq_comp_eq} indicially, as,
\begin{align}
 &\overline{c_i^{pl}}= c_i^{pl,*} + \chi^l_{ij}
\overline{\Delta \mu_j^{p}} - \zeta^l_i \overline{\Delta T^{p}}= \nonumber \\
&  c^\infty_i +  \Big[ I_0^* +
 \overline{\Delta \chi_{ij}^m} \overline{\Delta \mu_j^{m}} - \overline{\Delta \zeta_i^m} \overline{\Delta T^m} 
\nonumber \\
&  + \Delta c_i^{m,*} \Delta \eta_m\Big]  + \left[\left(\dfrac{\lambda}{{\eta_p^*} } 
- \dfrac{\lambda}{{\eta_p^*}^2} \Delta \eta_p\right) \tilde{f_i} \right],
\label{liq_comp_eq_2_cond}
\end{align}
where,
\begin{align}
 \tilde{f_i}=\dfrac{f_i^*}{l_i},
 \label{def_fipr}
\end{align}
and, $p$ is the index which represents a particular solid phase and is
not summed over in Eq.~\ref{liq_comp_eq_2_cond} as well as in the following equations. 
The quantities 
enclosed in square brackets in the RHS of Eq.~\ref{liq_comp_eq_2_cond} represent terms obtained by linearizing
the individual terms in the RHS of Eq.~\ref{liq_comp_eq}.
Eq.~\ref{liq_comp_eq_2_cond} can be re-written to express the diffusion potentials as a function
of the interfacial undercoolings and solid phase volume fractions
as,
\begin{align}
 &\left[\chi_{ij}^l - \overline{\Delta \chi_{ij}^p} \right]  \overline{\Delta \mu_j^{p}} 
  -\overline{\Delta \chi_{ij}^{m'}} \overline{\Delta \mu_j^{m'}} = \nonumber \\
  &\left(c_i^\infty+I_0^*+\dfrac{\lambda}{{\eta_p^*}}\tilde{f_i}
  -c_i^{pl,*}\right) \nonumber \\
  &-\overline{\Delta \zeta_i^{m'}} \overline{\Delta T^{m'}} - 
  \left( \overline{\Delta \zeta_i^{p}} - \zeta_i^l\right) \overline{\Delta T^p} \nonumber \\
  &+\Delta c_i^{m',*} \Delta \eta_{m'} + \left(\Delta c_i^{p,*}-\dfrac{\lambda}{{\eta_p^*}^2} \tilde{f_i}\right)\Delta \eta_p,
  \label{mu_func_eta_T_cond}
\end{align}
where, a summation over the index $m'$ runs from $1,\cdots,N$ leaving out $p$.  
So, Eq.~\ref{mu_func_eta_T_cond} represents a system of $N(K-1)$ equations which relates the $N(K-1)$ $\overline{\Delta \mu_i^{p}}$'s
to $\Delta \eta_p$'s and $\overline{\Delta T^p}$'s. To describe this dependence, we utilize the linearity of 
Eq.~\ref{mu_func_eta_T_cond} to write an explicit relation of the form,
\begin{align}
 \overline{\Delta \mu_i^{p}} = \overline{R_i^{p}} + R_i^{p, T^m} \overline{\Delta T^m} +  R_i^{p, \eta_m} \Delta \eta_m,
\label{mu_func_eta_T_cond_exp}
\end{align}
where, $\overline{R_i^{p}}$, $R_i^{p, T^m}$, and $ R_i^{p, \eta_m}$ are coefficients determined by solving Eq.~\ref{mu_func_eta_T_cond} with 
$m$ being the lone index which is summed over in Eq.~\ref{mu_func_eta_T_cond_exp}.
These relationships will enable the elimination of $\overline{\Delta \mu_i^{p l}}$
completely from the expressions of the undercoolings ahead of every solid phase given in Eq.~\ref{delT_ahead_of_p}.
Substituting Eq.~\ref{mu_func_eta_T_cond_exp} into Eq.~\ref{Taylor_I0_cond} we write in the indicial notation,
\begin{align}
 I_0&= I_0^*+ 
 \overline{\Delta \chi_{ij}^m} \left(\overline{R_j^{m}} + R_j^{m, T^v} \overline{\Delta T^v} +  R_j^{m, \eta_v} \Delta \eta_v\right)\nonumber \\
 &- \overline{\Delta \zeta_i^m} \overline{\Delta T^m}
 + \Delta c_i^{m,*} \Delta \eta_m,\nonumber \\
 &=\left(I_0^*+\overline{\Delta \chi_{ij}^m} \overline{R_j^{m}}\right) + 
 \left(\overline{\Delta \chi_{ij}^m} R_j^{m, T^v} \overline{\Delta T^v} - \overline{\Delta \zeta_i^m} \overline{\Delta T^m}\right)\nonumber \\
  &+ \left(\overline{\Delta \chi_{ij}^m} R_j^{m, \eta_v} \Delta \eta_v + \Delta c_i^{m,*} \Delta \eta_m\right),
 \label{I0_el_mu}
\end{align}
where $v$ is an index running over the phases $1,\cdots,N$ and is summed over along with the other phase index $m$.
The index $j$ denoting the components is also summed over while $i$ continues to represent a particular component.
Terms of similar character are collected within the parentheses in Eq.~\ref{I0_el_mu}. 

The resulting set of equations in~\ref{delT_ahead_of_p} ($N$ in number) 
are then solved for the $2N$ unknowns in $\overline{\Delta T^p}$ and $\Delta \eta_p$  
under the constraints of equality of undercoolings ahead of the solid phases given by Eq.~\ref{equal_undercool} ($N-1$ in number)
and the sum of the volume fractions of the solid phases adding up to unity in Eq.~\ref{sum_eta}.
Substituting Eq.~\ref{JH_avg_comp} into Eq.~\ref{delT_ahead_of_p}, we obtain, in Einstein's notation, without summing over $p$,
but summming over $i$,
\begin{align}
 \overline{\Delta T^p} &=  m_i^{p}\left(c_i^{l,*} - \overline{c_i^{pl}} \right) + \Gamma_p\overline{\kappa_p} \nonumber \\
 &= m_i^{p} \left[c_i^{l,*}-\left( c_i^\infty + I_0 + 
 \dfrac{\lambda}{\eta_p } \tilde{f_i} \right)\right] \nonumber \\
 &+\Gamma_p\overline{\kappa_p}.
 \label{delT_ahead_of_p2}
\end{align}
Linearizing the RHS of the above equation about equilibrium quantities and employing Eq.~\ref{I0_el_mu} leads to,
\begin{align}
  \overline{\Delta T^p} &= \Bigg[m_i^p \left(c_i^{l,*} - 
 \left(c_i^\infty + I_0^* + \overline{\Delta \chi_{ij}^{m}} \overline{R_j^m}+ \dfrac{\lambda}{{\eta_p^*}} \tilde{f_i}\right) \right)\nonumber \\
 &+ \Gamma_p \overline{\kappa_p^*}\Bigg] \nonumber \\
 &-\left[m_i^p \left(\overline{\Delta \chi_{ij}^{m}} R_j^{m, T^v}\overline{\Delta T^v} - \overline{\Delta \zeta_i^{m}} \overline{\Delta T^m}\right)\right] 
 \nonumber \\
 &-\Bigg[ m_i^p\Bigg(\overline{\Delta \chi_{ij}^{m}} R_j^{m, \eta_v}\Delta \eta_v +\Delta c_i^{m,*} \Delta \eta_{m}  \nonumber \\
 &-  \dfrac{\lambda}{{\eta_p^*}^2 } \tilde{f_i} \Delta \eta_p\Bigg)
 +\dfrac{\Gamma_p \overline{\kappa_p^*}}{\eta_p^*} \Delta \eta_p \Bigg],
  \label{JH_undercool_eta_dT}
\end{align}
where $\kappa_p^*$ is obtained by evaluating Eq.~\ref{kappa_def} for $\eta_p^*$. 
The three terms each enclosed in square brackets in the RHS of the above equation contain the
constants, and the terms linear in $\overline{\Delta T^m}$ and $\Delta \eta_m$ respectively.
We now impose Eq.~\ref{equal_undercool} on Eq.~\ref{JH_undercool_eta_dT} to re-express it in terms of
$\overline{\Delta T}$ and $\overline{\Delta \eta_p}$'s, as follows,
\begin{align}
 &\left[1+m_i^{p}\left( \overline{\Delta \chi_{ij}^{m}} \sum_{v=1}^N R_j^{m, T^v} -
 \sum_{m=1}^N\overline{\Delta \zeta_i^{m}} \right)\right]\overline{\Delta T} \nonumber \\
 &+ \Bigg[m_i^p \left(\overline{\Delta \chi_{ij}^{m}} R_j^{m, \eta_p} +\Delta c_i^{p,*} 
 -  \dfrac{\lambda}{{\eta_p^*}^2 } \tilde{f_i} \right) \nonumber \\
 &+\dfrac{\Gamma_p \overline{\kappa_p^*}}{\eta_p^*}\Bigg] \Delta \eta_p  \nonumber \\
 &+ \left[m_i^p\left(\overline{\Delta \chi_{ij}^{m}} R_j^{m, \eta_{v'}} +\Delta c_i^{v',*}\right) \right] \Delta \eta_{v'} =\nonumber \\
 &\Bigg[m_i^p \left(c_i^{l,*} - 
 \left(c_i^\infty + I_0^* + \overline{\Delta \chi_{ij}^{m}} \overline{R_j^m} + \dfrac{\lambda}{{\eta_p^*}} \tilde{f_i}\right)  \right)\nonumber \\
 &+ \Gamma_p \overline{\kappa_p^*}\Bigg],
 \label{JH_undercool_eta_eq_dT}
\end{align}
where, $v'$ is another phase index running from $1$ to $N$ except $p$ and is summed over along with the other phase index $m$.
The component indices $i$ and $j$ are also summed over in Eq.~\ref{JH_undercool_eta_eq_dT}. 
At this stage, we will invoke Eq.~\ref{sum_eta} to eliminate $\Delta \eta_N$ from the above set of equations.
Now, Eq.~\ref{JH_undercool_eta_eq_dT} represents a system of $N$ linear equations containing the same number of
unknowns in $\overline{\Delta T}$ and $\Delta \eta_1,\cdots,\Delta \eta_{N-1}$. Solving Eq.~\ref{JH_undercool_eta_eq_dT}
to compute $\overline{\Delta T}$ and $\Delta \eta_1,\cdots,\Delta \eta_{N-1}$, enables the 
calculation of $\overline{\Delta \mu_j^{1 }},\cdots,\overline{\Delta \mu_j^{N }}$
from Eq.~\ref{mu_func_eta_T_cond_exp}. Thereafter, the phase compositions can be directly obtained 
from Eqs.~\ref{gen_avg_sol_comp} and~\ref{gen_avg_liq_comp}.

To summarize, our analytical method involves the following steps:
\begin{itemize}
 \item For every solid-liquid equilibrium in a single period of the eutectic, the solid and the liquid compositions
 averaged over their corresponding lamellar widths are expressed as linear functions of the corresponding changes 
 in diffusion potentials and undercoolings.
 \item We also determine the average composition in the liquid in equilibrium with the different solid phases 
 from the Jackson-Hunt type analysis involving the superposition of multiple Fourier modes, 
 which is again linearized about the chosen eutectic point to obtain the liquid compositions 
 as functions of changes in diffusion potentials, undercoolings and changes in solid phase volume fractions.
 This also involves expressing the boundary layer compositions in terms of the departure of diffusion potentials and phase
 fractions from their corresponding values at the eutectic along with the undercooling at the solid-liquid interface.
 \item Using the equality of the liquid compositions obtained from the earlier steps we derive 
 expressions of the diffusion potentials as functions of undercoolings and changes in solid phase fractions.
 \item  Substituting for the liquid compositions using the linearized version of the Fourier series representation
 in terms of undercoolings, diffusion potentials and solid phase fractions  
 into the expressions of the undercoolings at each solid-liquid interface, and enforcing the isothermal nature of the interface we
 compute the magnitude of the interfacial undercooling and solid phase fractions.
 \item Phase compositions get automatically determined due to their explicit 
 and implicit (due to diffusion potentials) dependence on undercooling  
 alongside their dependence on solid phase fractions. 
\end{itemize}

A point to note over here is while Eq.~\ref{mu_func_eta_T_cond_exp} relates 
the deviations of the diffusion potentials $\overline{\Delta \mu_i^{p}}$
with the deviations of the solid fractions and the undercoolings, one can
additionally invoke the condition of local thermodynamic equilibrium at 
the interface (including curvature) and thereby eliminate the undercoolings
from the relation in Eq.~\ref{mu_func_eta_T_cond_exp}. A similar approach has
been used by Senninger and Voorhees \cite{Senninger2016}. We have tried
this out as well and the results from both approaches are comparable.
This completes our theoretical derivation of generic multi-component
multi-phase eutectic growth. In the following section, we validate our
theory against phase-field simulations of invariant and mono-variant
eutectic growth.
\section{Phase field model}
Following~\cite{Choudhury+11-3}, the grand potential functional($\Omega$) can be expressed as,
 \begin{align}
  {\Omega}\left(\vmu,T,\vphi\right)&=\int_{V}\Bigg[\Psi\left(\vmu,T,
 \vphi\right) \nonumber \\ 
  &+ \left(\epsilon a(\vphi,\nabla \vphi) +
 \dfrac{1}{\epsilon}w\left(\vphi\right)\right)\Bigg]dV,
  \label{GrandPotentialfunctional}
 \end{align}
 where $\vphi=\left[\phi_1,\phi_2, \cdots,\phi_N\right]$ are the phase-fields representing the 
 spatial arrangement of $N$ phases and \\
 $\vmu=\left[\mu_1, \mu_2,\cdots, \mu_{K-1}\right]$ are the diffusion potentials associated with 
 each one of the $K-1$ independent solutes.
 The functionals $w$ and $a$ represent the surface potential and the gradient energy density respectively. 
 The minimization of $\Omega$ leads to the evolution of the spatial arrangement of the phases ($\vphi$) denoted by,   
\begin{align}
\tau \epsilon \dfrac{\partial \phi_p}{\partial t}= 
\epsilon \left(\nabla \cdot \dfrac{\partial a(\vphi,\nabla \vphi)}{\partial \nabla \phi_p} - 
\dfrac{\partial a(\vphi,\nabla\vphi)}{\partial \phi_p} \right) \nonumber \\
-\dfrac{1}{\epsilon}\dfrac{\partial w\left(\vphi\right)}{\partial \phi_p} - \dfrac{\partial
\Psi\left(\vmu, T, \vphi\right)}{\partial \phi_p} - \Lambda,
\label{Equation6_grandchem}
\end{align}
where $\Lambda$ is calculated to ensure $\sum_{m=1}^N\phi_m=1$ at every 
mesh point in the simulation domain. $\tau$ is the relaxation constant with its value set 
based on the criterion stated in~\cite{Choudhury+11-3} and~\cite{Choudhury+12} to obtain a diffusion controlled
interface motion.

In this model, 
\begin{align}
\Delta \Psi^{mp}= \Psi^m-\Psi^p,
\label{driv_force}
\end{align}
represents the driving force for a transformation of phase $m$ to $p$, with the grand-potentials of the individual phases given by,
\begin{align}
 \Psi^{p}=f^{p}({\textbf{c}}^{p}\left(\vmu,T\right),T)-\sum_{i=1}^{K-1}\mu_i c_i^{p}\left(\vmu,T\right).
 \label{ind_gpot}
\end{align}
All the grand-potentials of the participating phases ($\Psi^p$)'s at any particular point in the simulation domain 
are interpolated to obtain $\Psi$ as,  
\begin{align}
 \Psi\left(\vmu,T,\phi\right) &= \sum^p\Psi^p (T, \vmu) h_p (\vphi),
 \label{GP_interpolation}
\end{align}
where,
\begin{align}
 h_p(\vphi)=\phi_p^2 \left(3- 2\phi_p \right) + 2 \phi_p \sum_{\substack{m=1,n=1, \\ m<n,m\neq p,n\neq p}}^{N,N} \phi_m\phi_n.
 \label{h_def}
\end{align}
The gradient energy density ($a(\vphi,\nabla \vphi)$) in the absence of interfacial energy anisotropy can be 
written as,
\begin{align}
 a(\vphi,\nabla \vphi)=\sum_{m=1,p=1,m < p}^{N,N} \gamma_{mp} {\big{|}q_{mp}\big{|}}^2,
 \label{a_def}
\end{align}
where, $\gamma_{mp}$ is the $m-p$ interfacial energy, and $q_{mp}$ is the normal vector to the $m-p$ interface, written as,
\begin{align}
 q_{mp}= \phi_m \nabla \phi_p -  \phi_p \nabla \phi_m.
 \label{q_def}
\end{align}
The surface potential $w(\phi)$ is given by,

\begin{align}
w\left(\vphi\right)=\left\{
\begin{array}{ll}
\dfrac{16}{\pi^{2}}\displaystyle \sum_{\substack{
m, p = 1 \\
(m < p)}}^{N, N}
\gamma_{m p}\phi_{m}\phi_{p}+ \nonumber \\ \displaystyle \sum_{\substack{
m, p, n= 1 \\
(m < p < n)}}^{N, N, N}
\gamma_{m p n}\phi_{m}\phi_{p}\phi_{n}, \quad
& \textrm{if} \hspace{0.1cm} \vphi \in \sum \vspace{0.5cm} \\
\infty, & \textrm{elsewhere}
\\
\end{array}
\right.
\end{align}
where $\sum=\{ \vphi \, | \sum_{m=1}^{N}\phi_{m} = 1$ and $\phi_{m} \geq 0 \}$,
$\gamma_{m p}$ is the surface entropy density and $\gamma_{m p n}$  
is a term added to maintain the solution at an $m p$ interface 
strictly along the two phase interface.

The evolution of $\vmu$ is expressed as,
\begin{align}
 &\left\lbrace\dfrac{\partial \mu_i}{\partial t}\right\rbrace = \nonumber\\
&\left[\sum_p 
h_p\left(\vphi\right)\dfrac{\partial c_i^p\left(\vmu,
T\right)}{\partial \mu_j}\right]^{-1}_{ij}\nonumber \\
&\Big\lbrace\nabla\cdot\left(\sum_{j=1}^{K-1}M_{ij}
\left(\vphi\right)\nabla \mu_j - \mathbf{J}_{at,i}\right)\nonumber\\ 
&- \sum_\nu c^p_{i}\left(\vmu,T\right)\dfrac{\partial
h_p\left(\vphi\right)}{\partial t}\Big\rbrace_j,
\label{Mu_explicit_temperature}
\end{align}
where $i$ and $j$ iterate over the $(K-1)$ independent components. $\left[\cdot\right]$ denotes 
a matrix of dimension $((K-1) \times (K-1))$  while $\left\lbrace \cdot \right\rbrace$ represents a vector 
of dimension $(K-1)$. The anti-trapping current $\mathbf{J_{at,i}}$ has a sense and magnitude which  
nullifies solute trapping at the solid-liquid interface and is determined by the expressions 
given in~\cite{Choudhury+12}.

The atomic mobility, $M_{ij}\left(\vphi\right)$ is 
obtained by interpolating the individual phase mobilities as,
\begin{align}
 M_{ij}\left( \vphi \right) &=\sum_\nu M_{ij}^p g_p(\vphi),
 \label{interp_mobility}
\end{align}
where the individual phase mobilities are given by,
\begin{align}
 \left[M_{ij}^p\right] &=  \left[D^p_{ik}\right] \left[\dfrac{\partial
c_k^p\left(\vmu,T\right)}{\partial \mu_j}\right],
\end{align}
where $D_{ij}^p$ are the solute inter-diffusivities in the $p$-th phase and $g_p(\vphi)$
are interpolants given as,
\begin{align}
 g_p(\vphi)=\phi_p^2 \left(3- 2\phi_p \right).
 \label{g_def}
\end{align}
The composition fields are obtained as functions of $\vmu$ and $\vphi$ as,
\begin{align}
 c_i &= \sum_p c_i^p\left(\vmu, T\right) h_p(\vphi), \nonumber \\ 
 c_i^p\left(\vmu, T\right) &=
-V_m\dfrac{\partial \Psi_p\left(\vphi,\vmu,T\right)}{\partial \mu_i}.
\label{c_of_mu}
\end{align}
with the molar $V_m$ is taken to be a constant across all the components.

\section{Results: Two-solid phases in a ternary system}
In this section we employ both the analytical and phase-field models described above
to study the solidification of two solid phases in a ternary alloy
and compare the predictions from these two techniques for 
different solute interdiffusivities in the liquid.
Although, the analytical theory and phase field model
are general enough to describe the solidification
at off-eutectic compositions, 
we restrict our study to eutectic compositions only. 

The solid phases in the ternary monovariant 
eutectic are anointed as $\alpha$ and $\beta$ 
with the independent components constituting the ternary alloy
being $A$ and $B$. All our studies are performed 
at a single sample pulling velocity of $V=0.01$ under an
imposed thermal gradient of $G=0.0005$,
with the solute diffusivities assumed to be
negligibly small in the solids compared to that in the liquid.     
We will first consider a model alloy whose solid compositions
are symmetrically located with respect to that 
of the liquid and follow it up with a similar study 
of a monovariant eutectic reaction in the Ni-Al-Zr alloy system.

\subsection{Calculation of $\eta_\alpha^*$ and $\eta_\beta^*$}
A three phase equilibrium ($\alpha$, $\beta$ and liquid) in a ternary system
is associated with a single degree of freedom as it can exist over a 
range of temperatures. During directional solidification, the far-field 
liquid composition can be found to correspond to a particular temperature ($T^*$)   
in the equilibrium phase diagram at which it is in equilibrium with two other 
solid phases. If such a liquid is assumed to solidify at this temperature,
the volume fractions of the resultant $\alpha$ and $\beta$ phases are 
what we refer to as $\eta_\alpha^*$ and $\eta_\beta^*$, respectively.
As there are two independent far-field compositions in a ternary system, we invoke
two artificial phase fractions $\eta_\alpha'$ and $\eta_\beta' \neq (1-\eta_\alpha')$
to solve for,
\begin{align}
 c_A^\infty=c_A^\alpha \eta_\alpha' + c_A^\beta \eta_\beta', \nonumber \\
 c_B^\infty=c_B^\alpha \eta_\alpha' + c_B^\beta \eta_\beta',
 \label{eta_det}
\end{align}
consistently. In general, $\eta_\alpha' + \eta_\beta' \neq 1$ and we compute 
the normalized volume fractions,
\begin{align}
 \eta_\alpha^*=\dfrac{\eta_\alpha'}{\eta_\alpha'+\eta_\beta'}, \nonumber \\
 \eta_\beta^*=\dfrac{\eta_\beta'}{\eta_\alpha'+\eta_\beta'},
 \label{eta_det_norm}
\end{align}
obeying $\eta_\alpha^* + \eta_\beta^* = 1$ and serving as values of the volume fractions
about which linearization is performed. 
 
\subsection{Symmetric system} 
To isolate and understand the effect of differences in solute 
diffusivities on the eutectic growth dynamics, we select 
a system  where the equilibrium phase compositions of solid phases are 
symmetric with respect to the liquid
composition ($c_A^\alpha=0.74$, $c_B^\alpha=0.18$, $c_A^\beta=0.18$, $c_B^\beta=0.74$, $c_A^l=0.36$, $c_B^l=0.36$).
Furthermore, we choose the liquidus slopes
to be consistent with this symmetry ($m_A^\alpha=0.45$, $m_B^\alpha=0$, $m_A^\beta=0$, $m_B^\beta=0.45$).
We begin with the study of a system with identical $\alpha$-liquid
and $\beta$-liquid interfacial energies which serves as a reference when we 
attempt to understand the dynamics of systems displaying dissimilar interfacial energies of the eutectic solids
with liquid.

\subsubsection{Equal $\alpha$-liquid and $\beta$-liquid interfacial energies}
\begin{figure}[!htbp]
\centering
\subfloat[]{\includegraphics[width=0.7\linewidth]{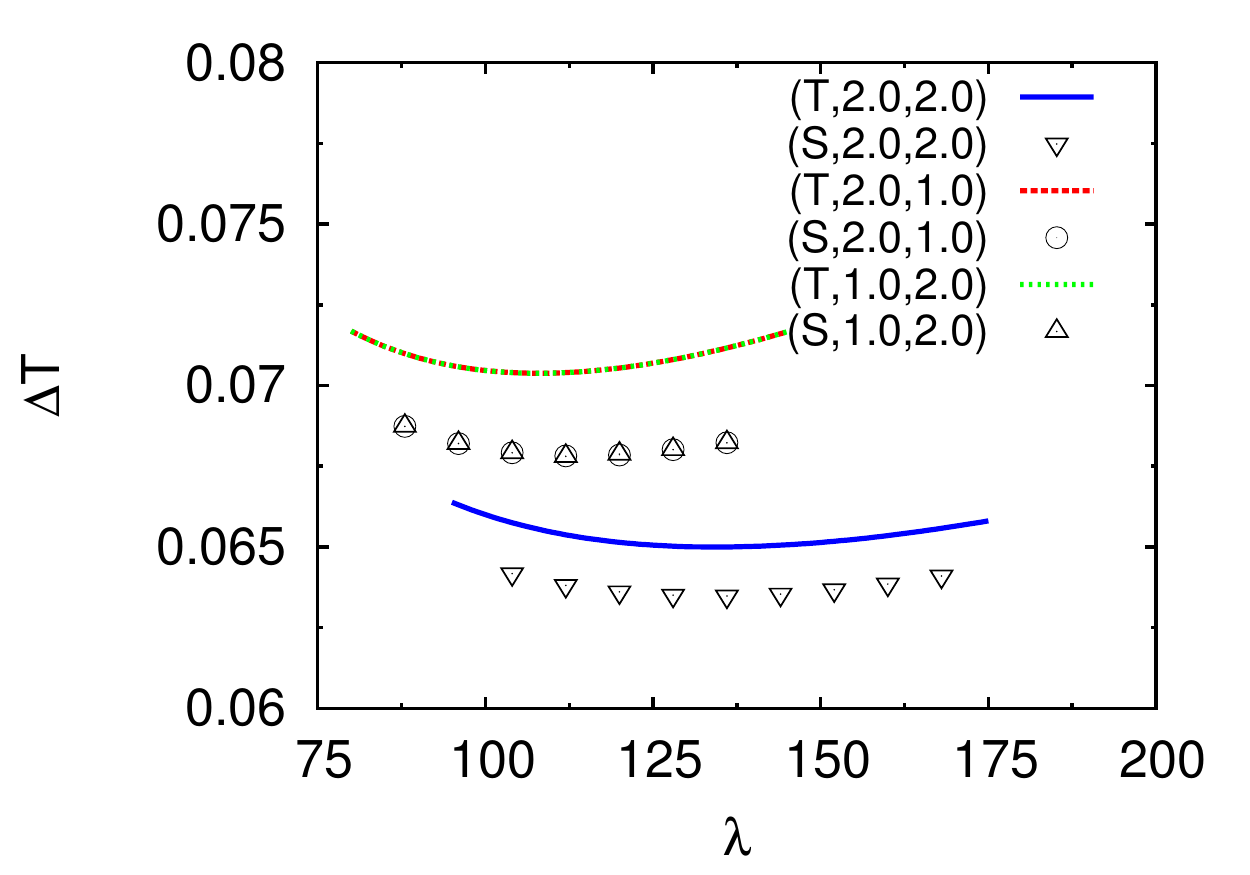}
\label{dtlam_3030}
}\,
\subfloat[]{\includegraphics[width=0.7\linewidth]{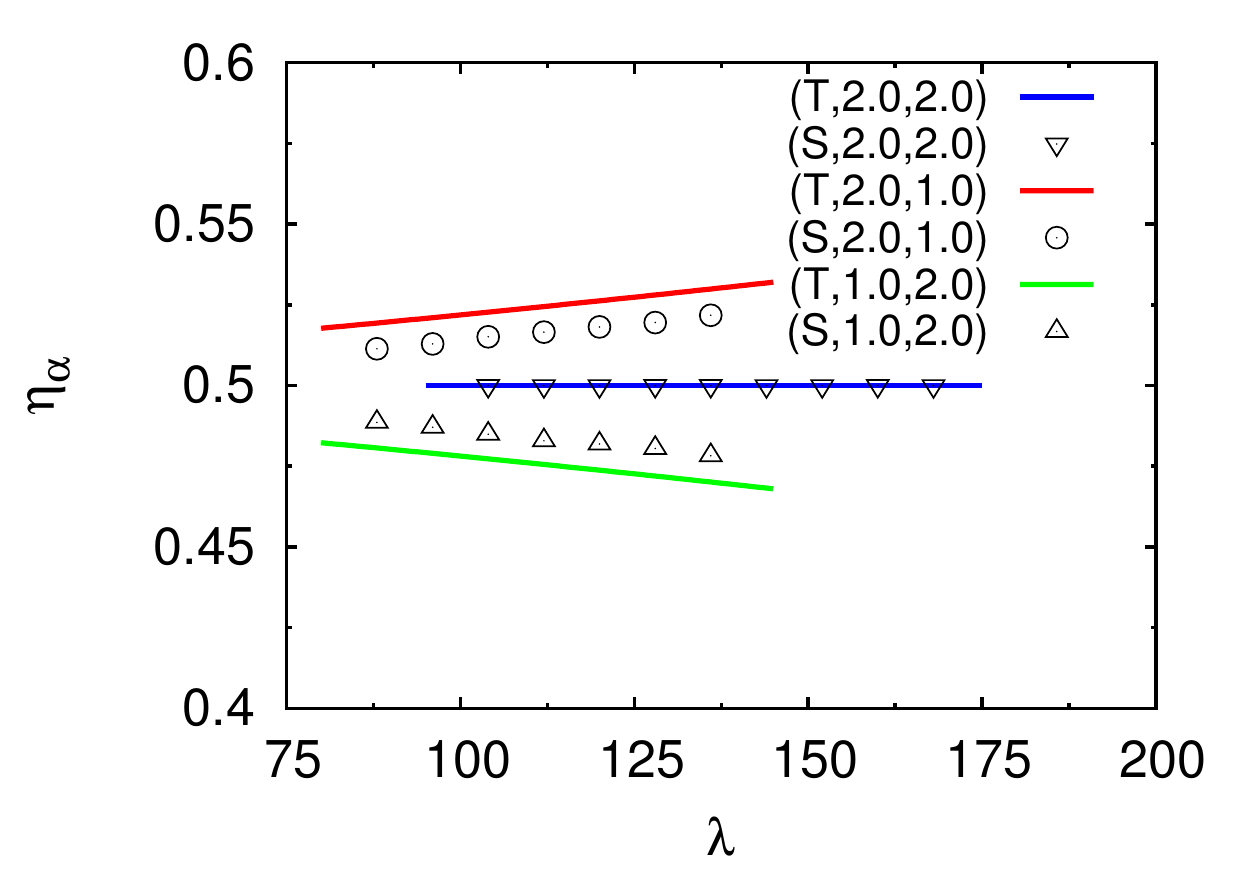}
\label{etlam_3030}
}\,
\subfloat[]{\includegraphics[width=0.7\linewidth]{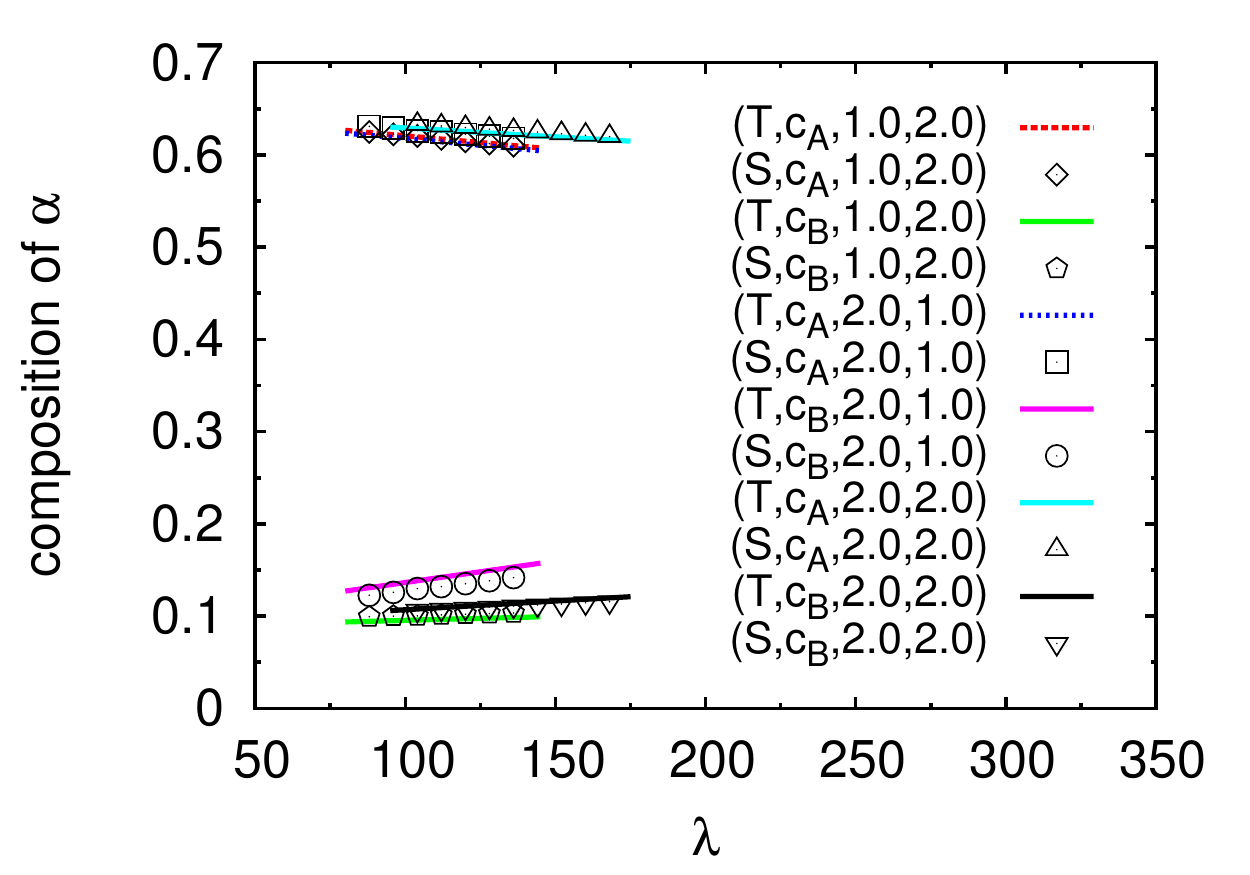}
\label{calp_3030}
}\,
\subfloat[]{\includegraphics[width=0.7\linewidth]{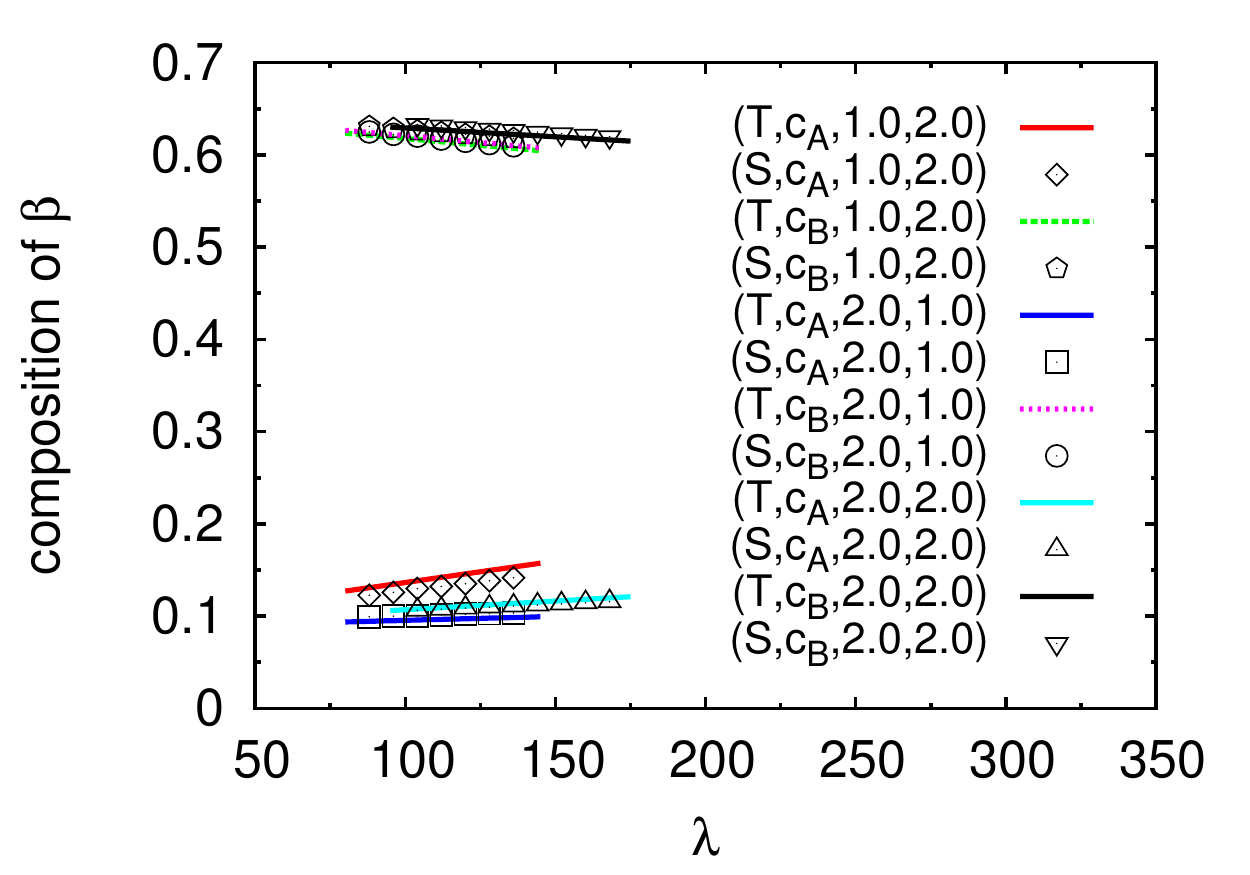}
\label{cbet_3030}
}
\caption{(Color online) Plots showing variations of (a)$\Delta T$, (b) $\eta_\alpha$, (c) $\alpha$ phase compositions
and (d) $\beta$ phase compositions, with $\lambda$, during two phase growth in a model symmetric ternary alloy. 
The first position in the figure legends in (a) and (b) 
indicates whether the plotted data comes from theory (T) or simulations (S); the second and third positions 
represent values of $D_{AA}$ and $D_{BB}$ respectively. For (c) and (d), the first position in the figure 
legends conveys the same as in (a) and (b), the second position indicates whether the
plotted variation in compositions corresponds to 
$A$ or $B$. The third and fourth positions in the figure legends in (c) and (d) represent the 
values of solutal interdiffusivities of $D_{AA}$ and $D_{BB}$ respectively.
}
\label{symm_3030}
\end{figure}

The equality of $\alpha$-liquid and $\beta$-liquid interfacial energies 
leads to $\theta_{\alpha\beta}=\theta_{\beta\alpha}=30\degree$ with the Gibbs-Thomson 
coefficients computed to be $\Gamma_\alpha=\Gamma_\beta=0.77$ for the particular thermodynamics employed. 
We present analytically calculated variations of interfacial undercoolings ($\Delta T$),
solid phase volume fractions ($\eta_\alpha$) and compositions of the $\alpha$ and the $\beta$
phases with lamellar width($\lambda$) in Fig.~\ref{symm_3030} and compare them against predictions 
obtained from phase field simulations. 
As can be seen from Fig.~\ref{dtlam_3030},
the analytical and phase-field calculations are in very good agreement as far as the 
predictions in $\lambda_{min}$ (the lamellar width corresponding to the minimum in $\Delta T$) 
are concerned. The close agreement between the analytical theory and the phase field simulations are also 
evident from the variations of $\eta_\alpha$ versus $\lambda$ presented in Fig.~\ref{etlam_3030}
and the variation of the average solid phase compositions in Figs.~\ref{calp_3030} and~\ref{cbet_3030}. 
The deviations of analytically computed values of $\Delta T$ 
(in Fig.~\ref{dtlam_3030}) from those obtained from phase field simulations
can be attributed to the inherent assumptions in the Jackson-Hunt calculations, where a 
planar interface is used to approximate the diffusion-field ahead of the solid interfaces,
which are in reality curved for a system with isotropic surface energies. This mismatch 
has also been shown before~\cite{Folch2005}, and a concomitant comparison of the phase-field 
method with calculations based on the boundary-integral method 
have proved that the predictions of the 
phase-field method are more accurate in this regard.

From Fig.~\ref{dtlam_3030}, it can be seen that lowering either of the solute diffusivities 
leads to a reduction in the length scale of the eutectic ($\lambda_{min}$) with a consequent
rise in $\Delta T$. This is a result of a lowered effective diffusivity leading to a 
lower effective diffusion length.
As a reflection of the inherent symmetry in the system, the theoretical calculations for $\Delta T$ vs $\lambda$ 
are exactly identical for $D_{AA}=2.0, D_{BB}=1.0$ and $D_{AA}=1.0, D_{BB}=2.0$. 

For equal solute diffusivities the volume fractions of the eutectic solids are the same, but Fig.~\ref{etlam_3030}
reveals that for $D_{AA}=2.0, D_{BB}=1.0$, $\alpha$ phase occupies a larger 
solid fraction ($\eta_\alpha>0.5$) of the lamellar width ($\lambda$); 
the same observation being valid for $\beta$ when $D_{AA}=1.0, D_{BB}=2.0$.


Physically, the change in the volume fractions can be seen as a consequence of
asymmetric changes in the constitutional undercooling at each solid-liquid interface.
Starting from a purely symmetric state with equal volume fractions, a situation 
of higher $D_{AA}$ in comparison to $D_{BB}$ would result
in a lower undercooling ahead of the $\alpha-l$ interface than the $\beta-l$
interface. To recover an isothermal situation between the two interfaces, would require
the $\alpha-l$ interface to assume an interfacial curvature that is greater
than that acquired by the $\alpha-l$ interface when mechanical equilibrium 
is maintained at the trijunction. This departure from equilibrium acts as
a driving force, where mechanical equilibrium is re-established through an 
increase in the volume fractions of the phase $\alpha$.
This explains the observed variation of $\eta_\alpha$ with
$\lambda$  in Fig.~\ref{etlam_3030}. A similar argument can be made with respect
to the lowering of the value of $D_{AA}$ with respect to $D_{BB}$ (see Fig.~\ref{etlam_3030}), 
where it must be noted that as a consequence of the underlying symmetry in the system  that the
$\eta_\alpha$ vs $\lambda$ curve for $D_{AA}=1.0, D_{BB}=2.0$ can be reflected about the
$\eta_\alpha=0.5$ line to obtain the variation for $D_{AA}=2.0, D_{BB}=1.0$.
In general, any change(change in diffusivity, interfacial energies etc.) 
which causes an increase in the undercooling of
a particular phase-liquid interface would need to be offset through 
an appropriate decrease in the interfacial curvature which can be 
affected only through a departure of the angles at the triple-point 
to lower values than that exists at equilibrium, keeping the same 
phase fractions. This departure acts
as a driving force to establish equilibrium which is achieved by 
a decrease in the volume fraction of this phase.

\subsubsection{Unequal $\alpha$-liquid and $\beta$-liquid interfacial energies}
\begin{figure}[!htbp]
\centering
\subfloat[]{\includegraphics[width=0.7\linewidth]{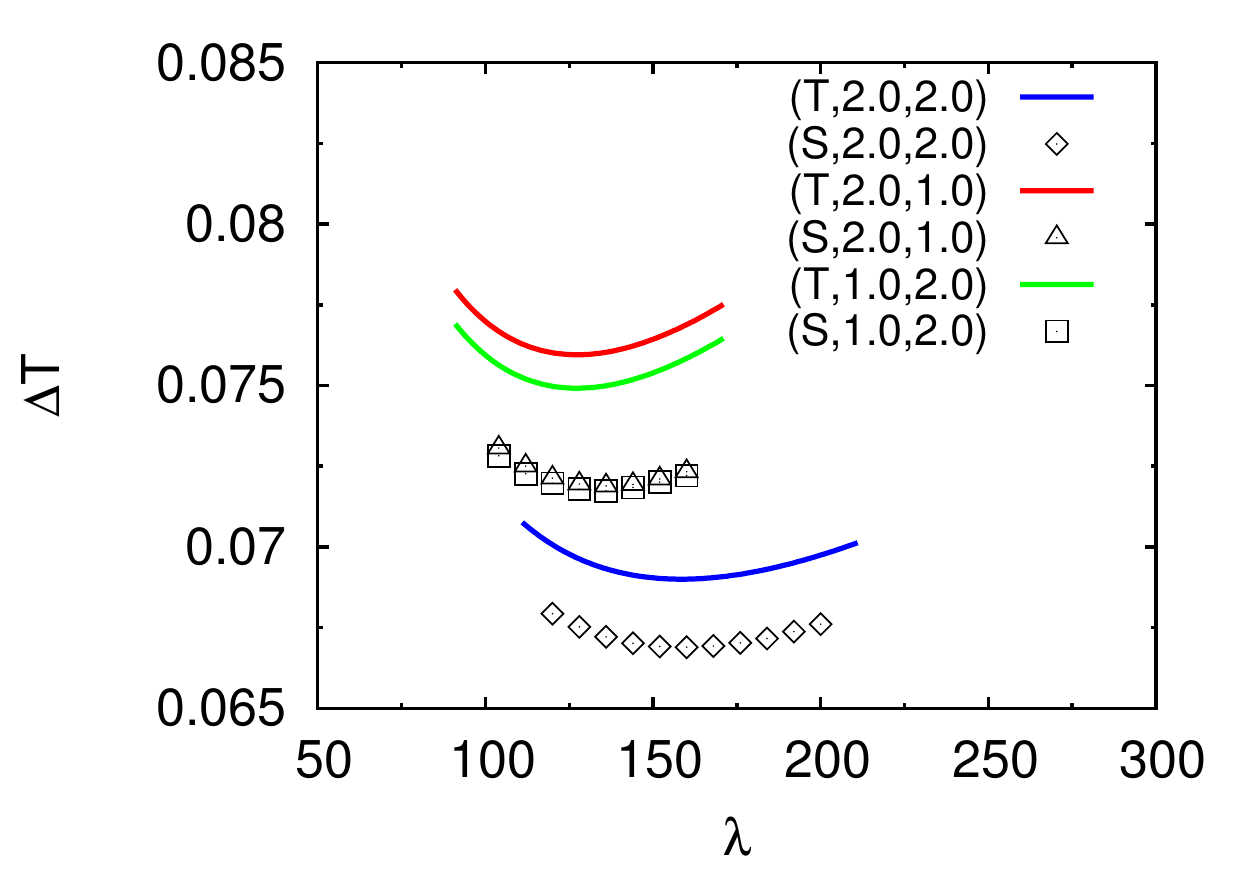}
\label{dtlam_3045}
}\,
\subfloat[]{\includegraphics[width=0.7\linewidth]{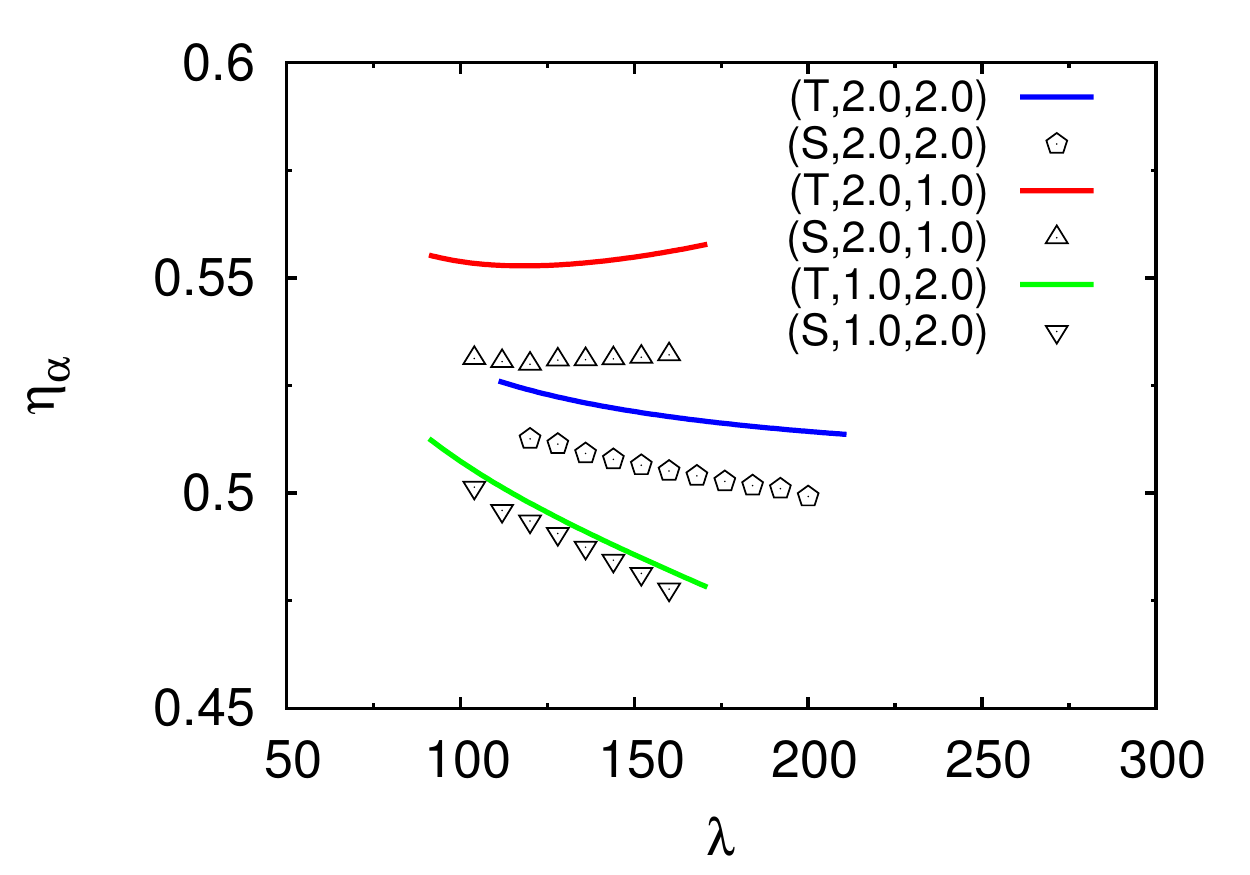}
\label{etlam_3045}
}\,
\subfloat[]{\includegraphics[width=0.7\linewidth]{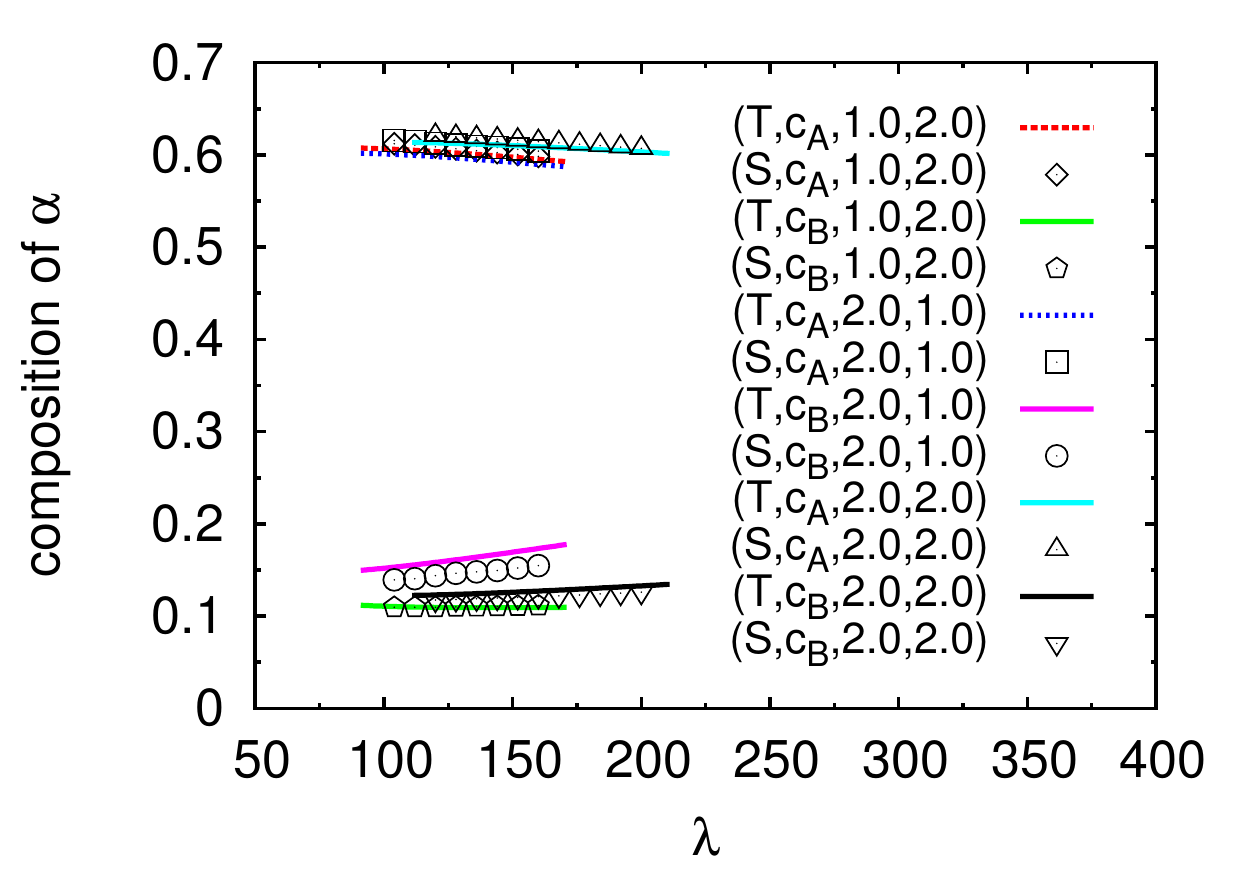}
\label{calp_3045}
}\,
\subfloat[]{\includegraphics[width=0.7\linewidth]{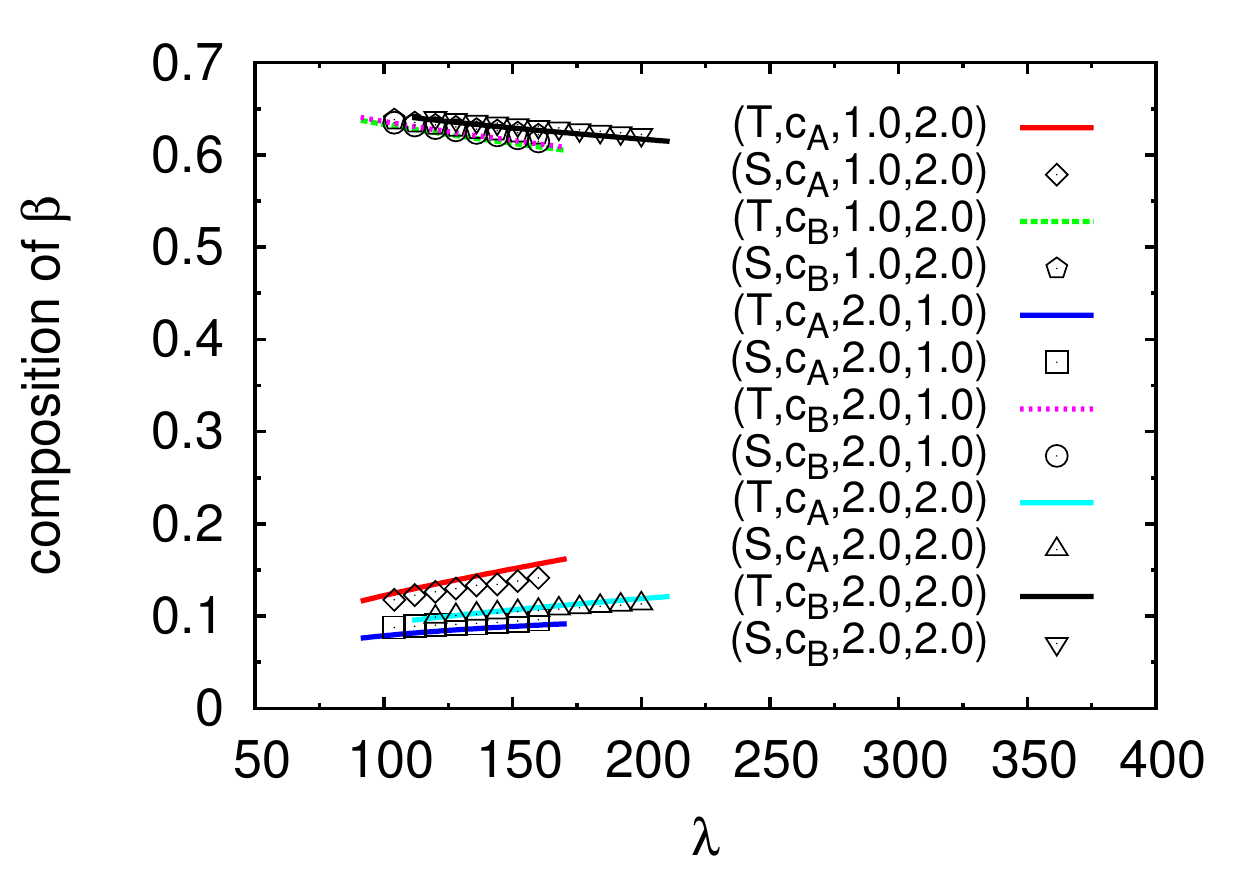}
\label{cbet_3045}
}
\caption{(Color online) Plots showing variations of (a)$\Delta T$, (b) $\eta_\alpha$, (c) $\alpha$ 
phase compositions, and (d)$\beta$ phase compositions,
with $\lambda$, during two phase eutectic growth in a model symmetric ternary alloy with unequal interfacial energies. 
The figure legends can be interpreted in the same way as described in the caption of
Fig.~\ref{symm_3030}.     
}
\label{symm_3045}
\end{figure}

Here we probe the effect of unequal surface energies on the steady-state monovariant eutectic 
growth dynamics while retaining the symmetry in the phase compositions and the 
liquidus slopes from the system just discussed.
The interfacial energies are so chosen such that $\theta_{\alpha\beta}=30\degree$
and $\theta_{\beta\alpha}=45\degree$ with the Gibbs-Thomson coefficients being $\Gamma_\alpha=0.77$
and $\Gamma_\beta=0.94$. Fig.~\ref{symm_3045} shows the variation in $\Delta T$, $\eta_\alpha$ and the 
solid phase compositions with $\lambda$ obtained from both analytical and phase field calculations.

The variation of $\Delta T$ with $\lambda$ reported in Fig.~\ref{dtlam_3045} presents a departure from the symmetry
observed in Fig.~\ref{dtlam_3030} as the curves corresponding to $D_{AA}=2.0, D_{BB}=1.0$ and $D_{AA}=1.0, D_{BB}=2.0$
do not overlap. This is a consequence of the dissimilar energies of the two eutectic solid-liquid interfaces.
However, the variation of the phase compositions, minimum undercooling spacings and 
the variation of the undercoolings are similarly captured by the simulations
and the theoretical calculations.

\begin{figure}[!htbp]
\centering
\includegraphics[width=0.7\linewidth]{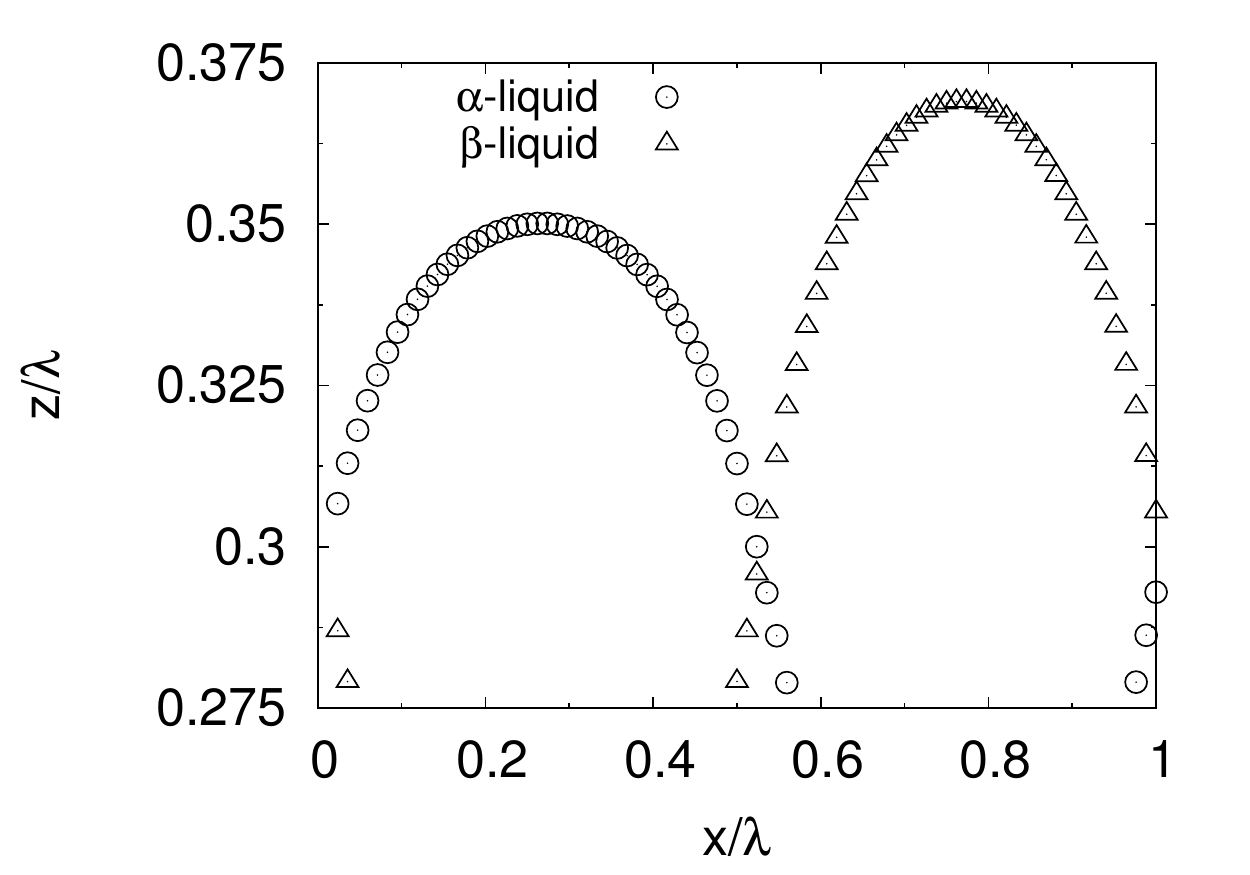}
\caption{Plot showing the locations of the $\alpha-l$ and the $\beta-l$ interfaces.}     
\label{int_shape}
\end{figure}

Furthermore, while the trends in the variation of the $\eta_{\alpha}$ with $\lambda$ are effectively
captured for all the cases, the magnitude of variation between the simulations and theory
is larger than the previous simulations with the symmetric interface properties. 
The reason for this is the asymmetric nature of the interface shapes, where
in the phase-field simulation the $\beta-l$ interface is ahead of the $\alpha-l$
interface (see Fig.\ref{int_shape}) and thereby clearly the interfacial undercoolings of the two phases
are not the same. Additionally, the departure from a planar interface is higher
for the $\beta-l$ interface compared to the $\alpha-l$ interface, which also 
implies that this brings in added asymmetry with respect to a mismatch with the
analytical calculations which are performed for a planar interface. Thereby, 
now any change in the interface shape which reduces the curvature of the 
$\beta-l$ interface decreases the disparity between the analytical calculations
and phase-field simulations and additionally with increasing curvature 
differences between the $\alpha-l$ and $\beta-l$ interfaces, the 
discrepancies between the theoretical predictions and simulation 
results also increases. Thus, this brings to light a limitation of
the analytical calculations, which work best when interfacial shapes
of the solid-liquid interfaces are similar.

\subsection{Ni-Al-Zr alloy system}
\begin{figure}[!htbp]
\centering
\subfloat[]{\includegraphics[width=0.7\linewidth]{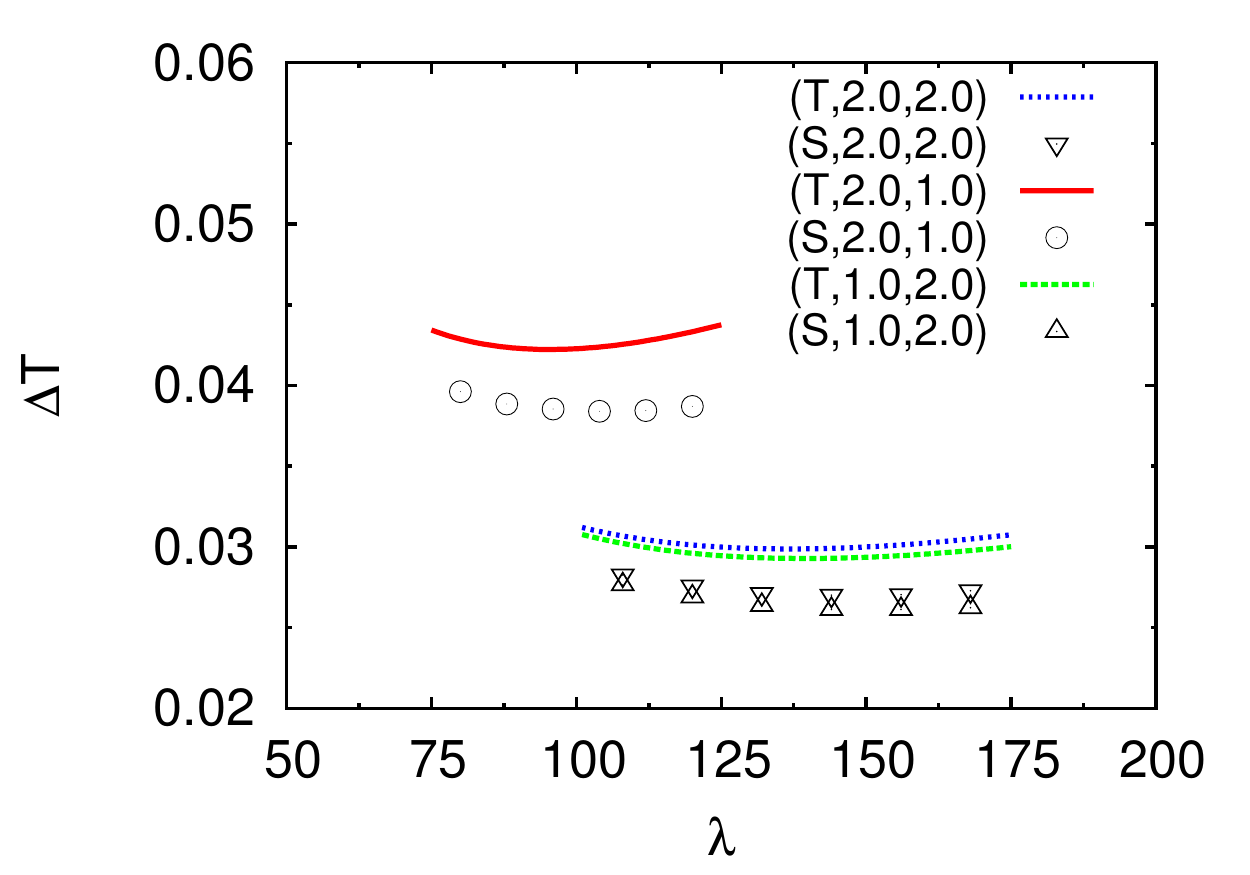}
\label{dtlam_asymm}
}\,
\subfloat[]{\includegraphics[width=0.7\linewidth]{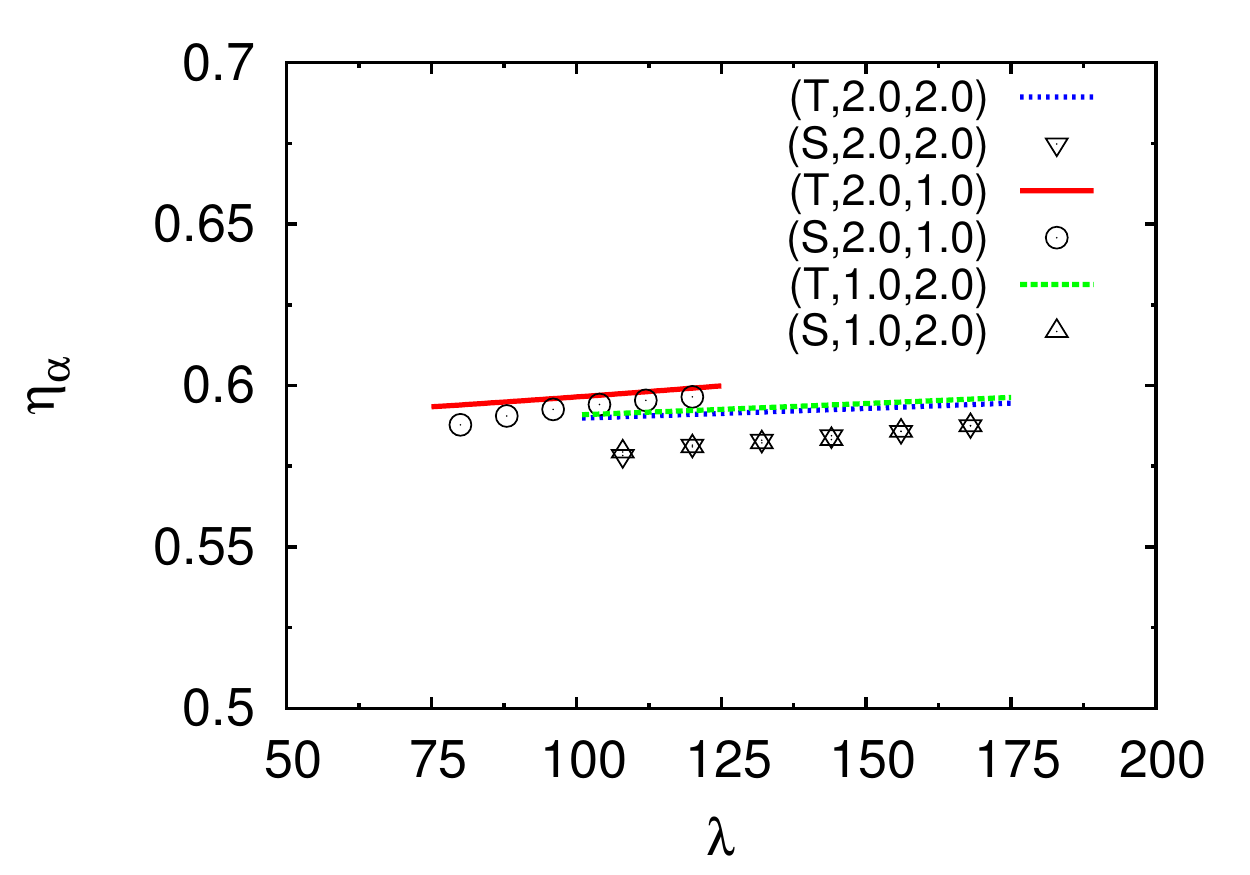}
\label{etlam_asymm}
}\,
\subfloat[]{\includegraphics[width=0.7\linewidth]{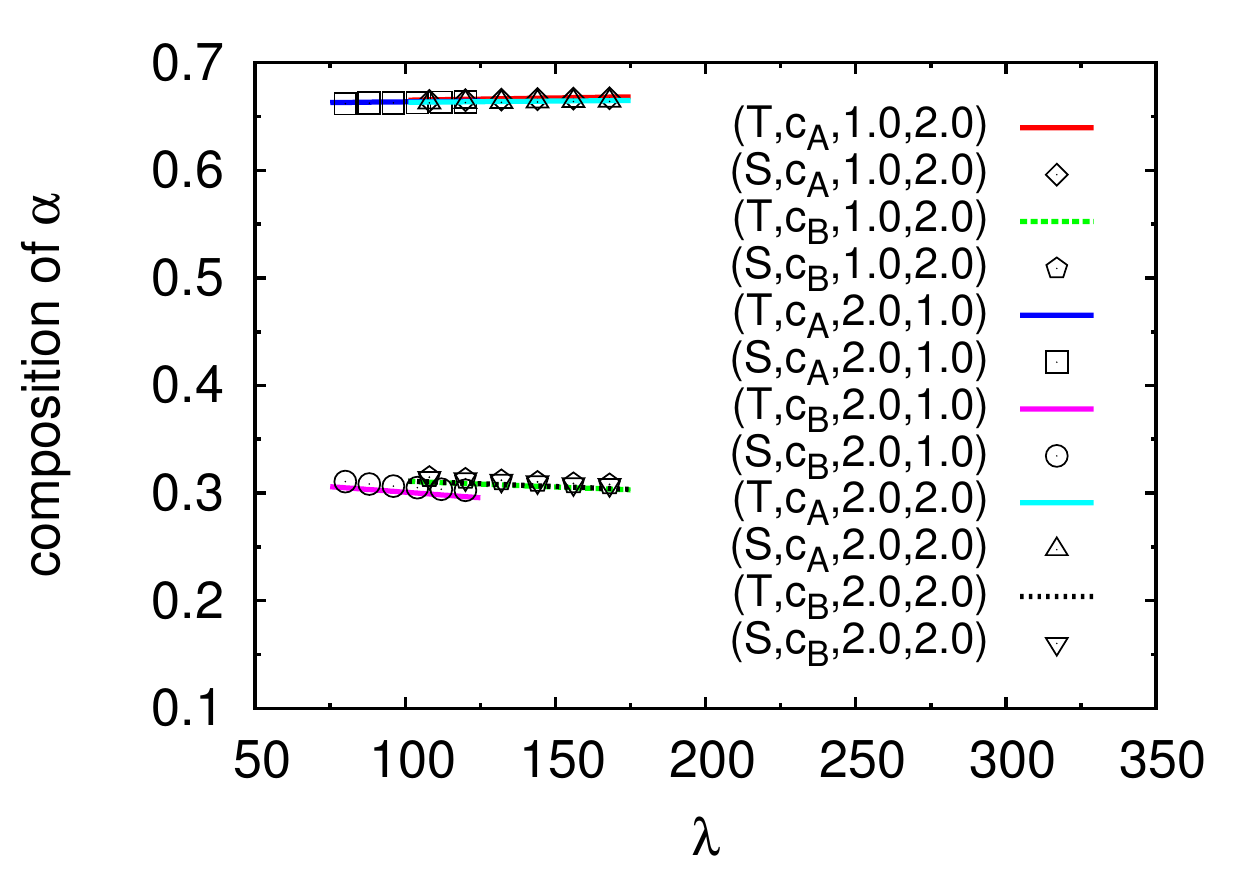}
\label{calp_asymm}
}\,
\subfloat[]{\includegraphics[width=0.7\linewidth]{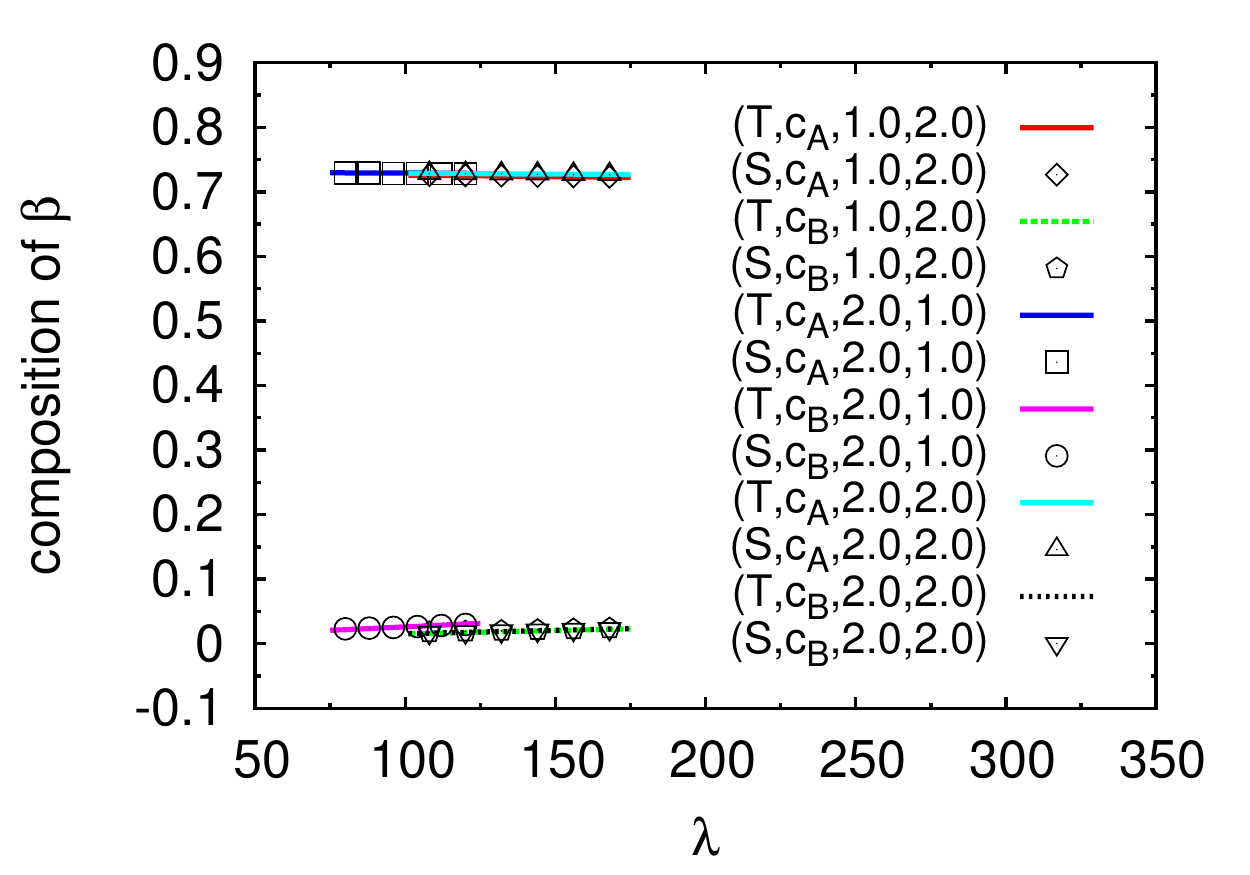}
\label{cbet_asymm}
}
\caption{(Color online) Plots showing variations of (a)$\Delta T$, (b) $\eta_\alpha$, (c) $\alpha$ phase compositions,
and (d)$\beta$ phase compositions,
with $\lambda$, during two-phase growth in a Ni-Al-Zr alloy.
The figure legends can be interpreted in the same way as described in the caption of
Fig.~\ref{symm_3030}.     
}
\label{asymm}
\end{figure}

In this section we study the steady-state dynamics of monovariant eutectic growth in a Ni-Al-Zr alloy at
the backdrop of the insights developed in the previous section. 
The equilibrium phase compositions at the temperature of interest are: 
$c_A^\alpha=0.67, c_B^\alpha=0.32, c_A^\beta=0.74, c_B^\beta=0.0034, c_A^l=0.69, c_B^l=0.19$ with the 
liquidus slopes being $m_A^\alpha=0.37, m_B^\alpha=1.08, m_A^\beta=-0.07, m_B^\beta=-1.0$. The Gibbs-Thomson 
coefficients are $\Gamma_\alpha=1.13,\Gamma_\beta=0.81$ with the contact angles $\theta_{\alpha\beta}=\theta_{\beta\alpha}=30\degree$.
We present the variations in $\Delta T$, $\eta_\alpha$ and the solid phase 
compositions in Fig.~\ref{asymm}.

From all the diffusivity combinations studied it can be said that 
the equilibrium phase compositions and the liquidus slopes are such that a
higher volume fraction of $\alpha$ is the preferred morphology.
Furthermore, it can be seen from Figs.~\ref{dtlam_asymm} and~\ref{etlam_asymm}, that the dynamics is much more
sensitive to a change in $D_{BB}$ compared to a change in $D_{AA}$. The undercoolings are found to be much 
higher for $D_{AA}=2.0, D_{BB}=1.0$ than for the other two situations studied, 
with an accompanied shift in length scales ($\lambda_{min}$) towards smaller values.
Also, $\eta_\alpha$ displays a steady rise with $\lambda$ which is much steeper for $D_{AA}=2.0, D_{BB}=1.0$ 
compared to the other two diffusivity configurations studied. The variations of the solid phase compositions
are depicted in Fig.~\ref{calp_asymm} and~\ref{cbet_asymm}.

\section{Results: Three phases in a ternary system}

Following up from the previous studies on two-phase
mono variant growth, in this section we investigate
three-phase invariant growth. Contrary to two-phase 
growth where there is a single possibility for the 
lamellar arrangement of the phases, for the case
of three-phase growth there exist infinitely many
configurations (e.g., $\alpha\beta\gamma, \alpha\beta\alpha\gamma \ldots$). 
In the following discussion,  
we consider two such possibilities for study, using 
both analytical calculations and phase-field
simulations. Here  we conduct simulations for the different
choices of the diffusivity matrices and compare
the predictions of the phase compositions and the 
volume fractions between the phase-field simulations
and the theoretical predictions.
The equilibrium phase compositions 
at the temperature of the invariant eutectic are: 
$c_A^\alpha=0.706, c_B^\alpha=0.146, c_A^\beta=0.146, c_B^\beta=0.706, c_A^l=0.333, c_B^l=0.333$ with the 
liquidus slopes being $m_A^\alpha=0.91, m_B^\alpha=0.0, m_A^\beta=0.0, m_B^\beta=0.91$, 
$m_A^\gamma=-0.91, m_B^\gamma=-0.91$. The Gibbs-Thomson 
coefficients are $\Gamma_\alpha=\Gamma_\beta=\Gamma_\gamma=1.558$
with the contact angles $\theta_{\alpha\beta}=\theta_{\beta\alpha}=\theta_{\beta\gamma}=\theta_{\gamma\beta}
=\theta_{\alpha\gamma}=\theta_{\gamma\alpha}=30\degree$. The directional solidification conditions
are kept the same as in the study of monovariant eutectic growth in ternary alloys. 

We first consider the
simplest arrangement $\alpha\beta\gamma$, where 
for the case of equal diagonal diffusivities, 
we get excellent agreement between our theory
and phase-field simulation results (see Fig. \ref{abg}), 
which is reflected not only in the variations of
the undercooling with spacing, but also in the compositions of the phases (see Fig.~\ref{abg_contd})
and in the volume fractions which given the symmetry of the 
phase-diagram and the diffusivities remain at 
($\eta_\alpha,\eta_\beta,\eta_\gamma$):(1/3,1/3,1/3).

\begin{figure}[!htbp]
\centering
\subfloat[]{\includegraphics[width=0.7\linewidth]{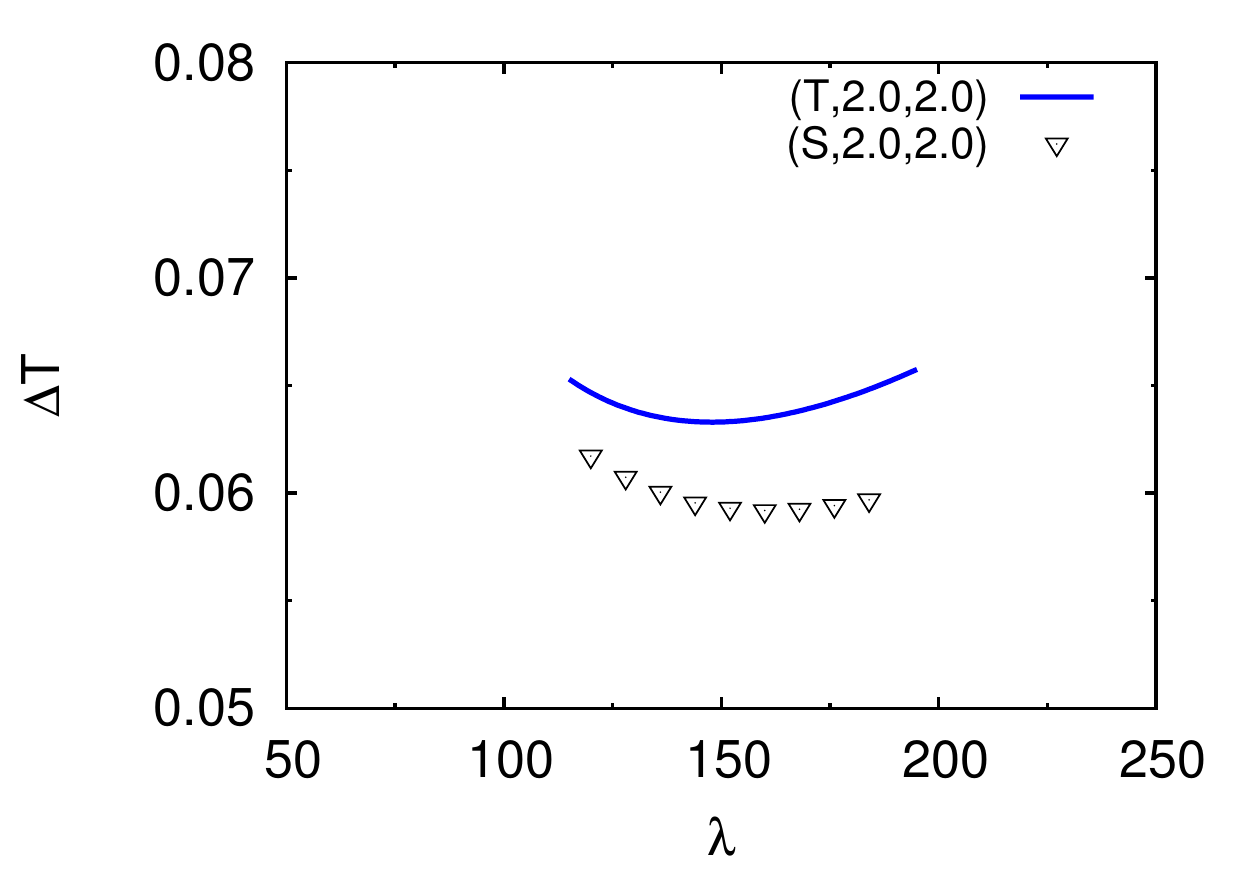}
\label{dtlam_abg}
}\,
\subfloat[]{\includegraphics[width=0.7\linewidth]{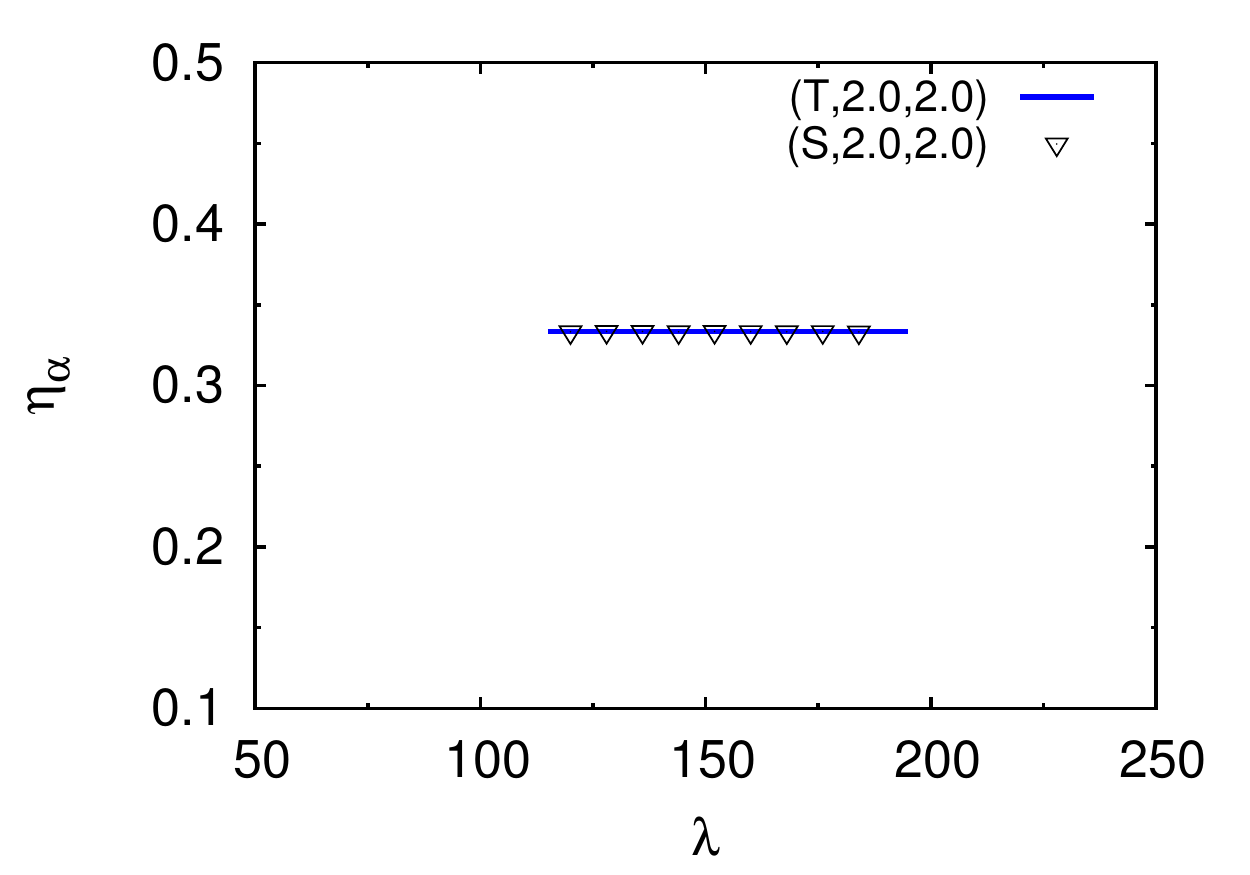}
\label{et_alp_lam_abg}
}\,
\subfloat[]{\includegraphics[width=0.7\linewidth]{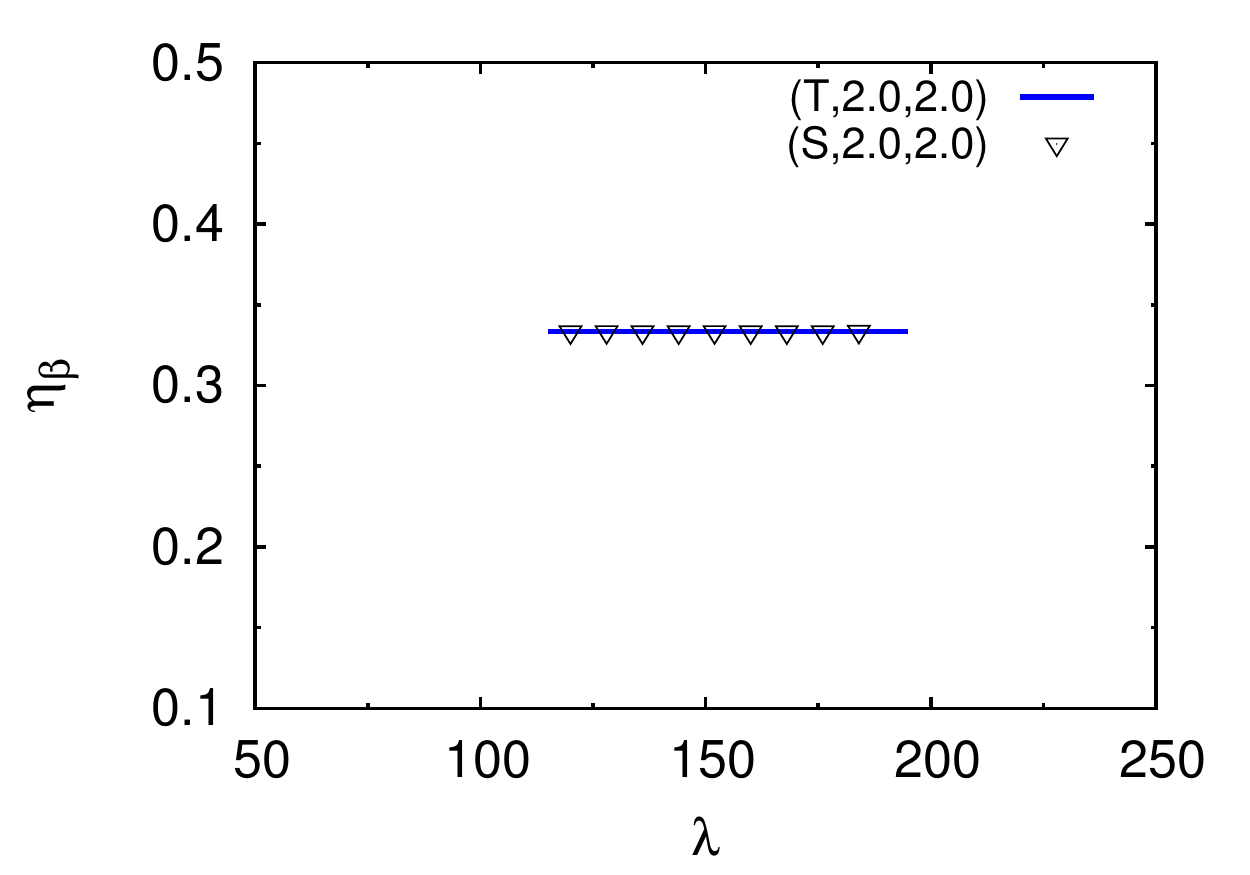}
\label{et_bet_lam_abg}
}
\caption{(Color online) Plots showing variations of (a)$\Delta T$, (b) $\eta_\alpha$, and (c) $\eta_\beta$,
with $\lambda$, during three phase eutectic growth in a model symmetric ternary alloy. A single
 wavelength of the eutectic solids has the configuration: $\alpha\beta\gamma$. 
The figure legends can be interpreted in the same way as described in the caption of
Fig.~\ref{symm_3030}.     
}
\label{abg}
\end{figure}

\begin{figure}[!htbp]
\centering
\ContinuedFloat
\subfloat[]{\includegraphics[width=0.7\linewidth]{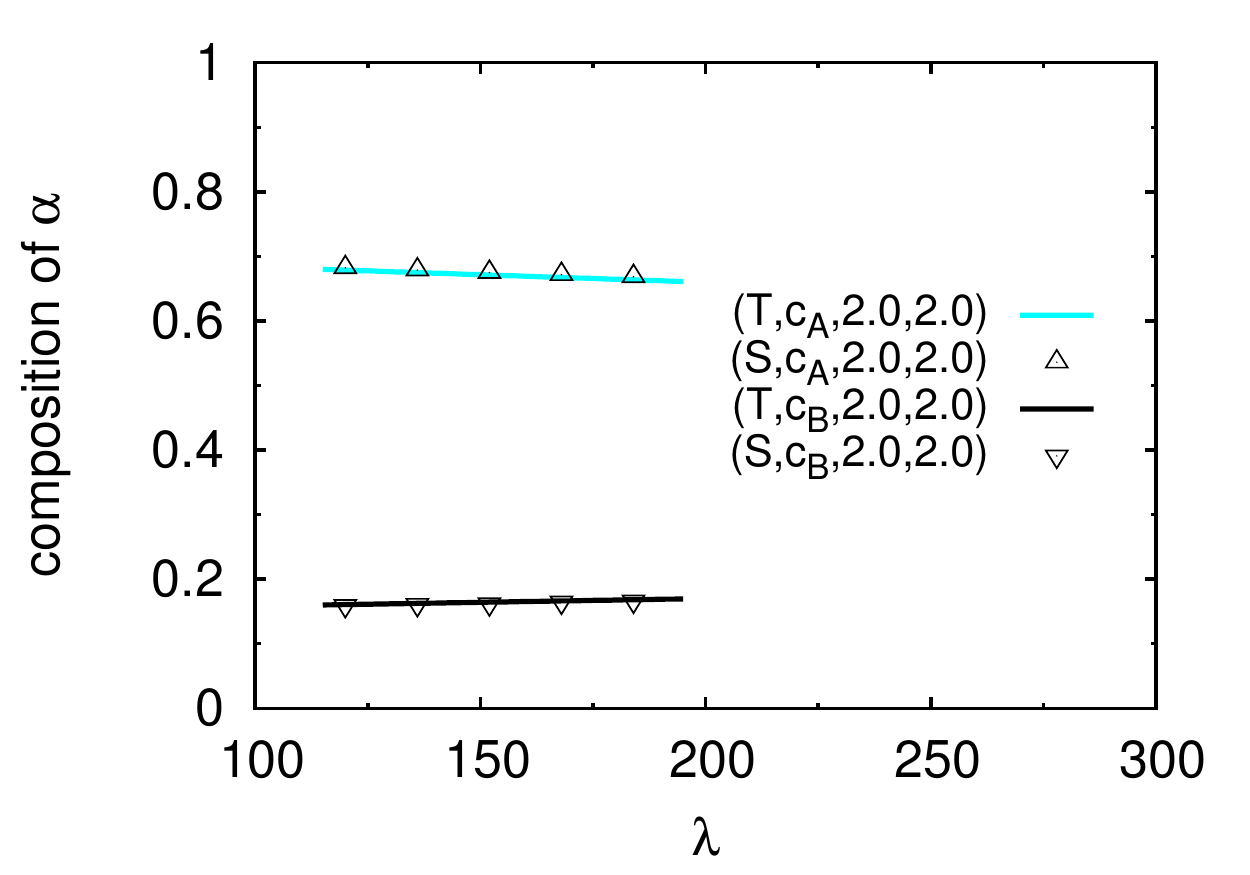}
\label{calp_abg}
}\,
\subfloat[]{\includegraphics[width=0.7\linewidth]{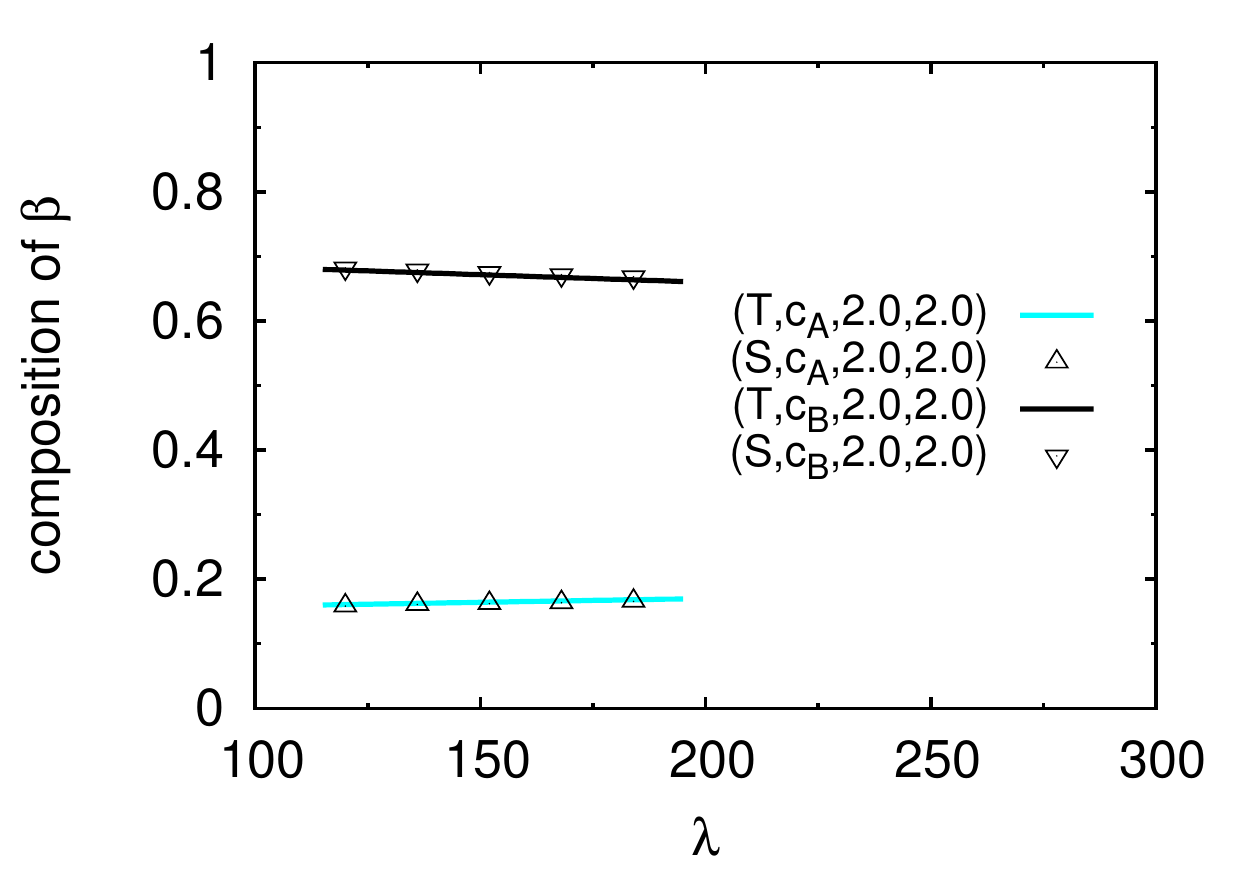}
\label{cbet_abg}
}\,
\subfloat[]{\includegraphics[width=0.7\linewidth]{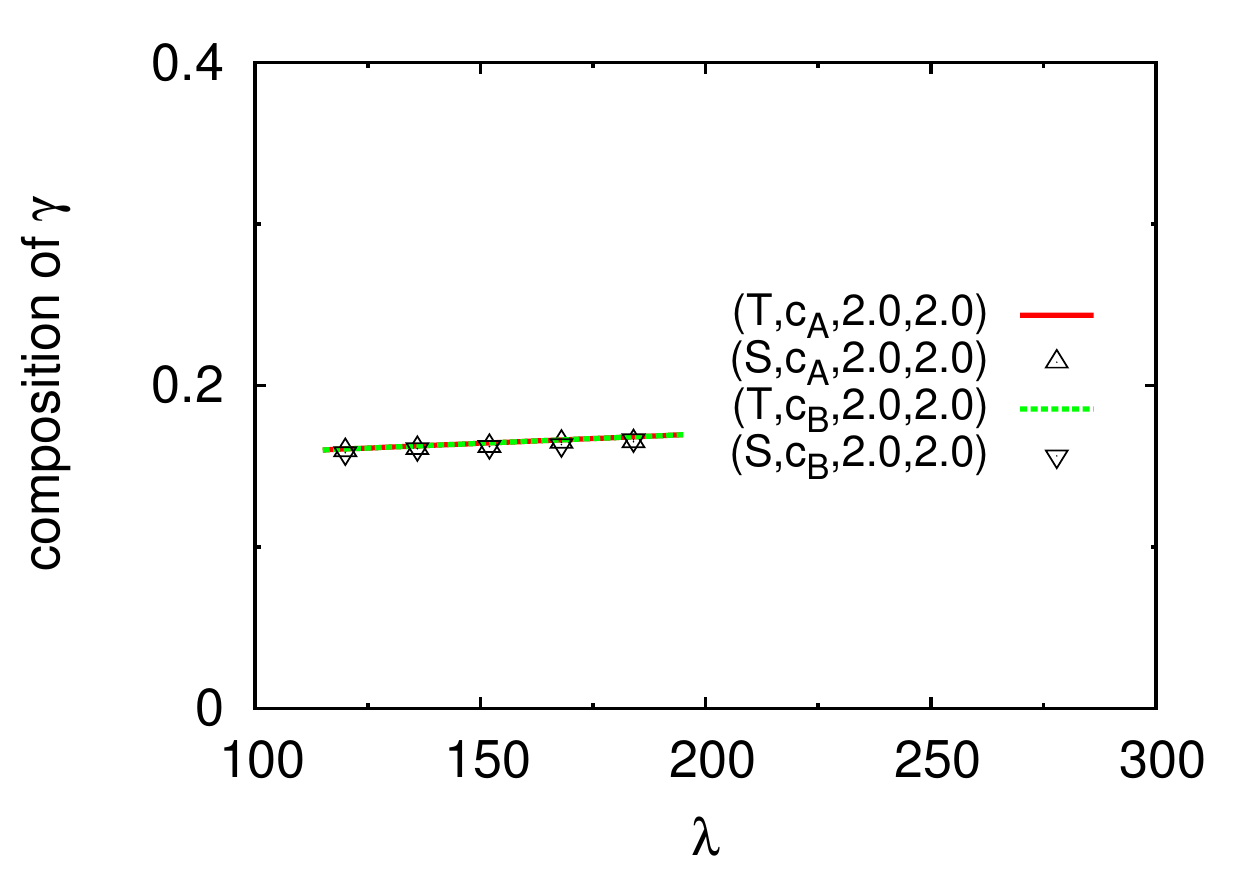}
\label{cgam_abg}
}
\caption{(Color online) Plots showing variations of (d)$\alpha$, (e)$\beta$, and (f)$\gamma$
phase compositions with $\lambda$, during three phase eutectic growth in a model symmetric ternary alloy. A single
 wavelength of the eutectic solids has the configuration: $\alpha\beta\gamma$. The figure legends can be interpreted in 
the same way as described in the caption of
Fig.~\ref{symm_3030}.     
}
\label{abg_contd}
\end{figure}

\begin{figure}[!htbp]
\centering
\subfloat[]{\includegraphics[width=0.7\linewidth]{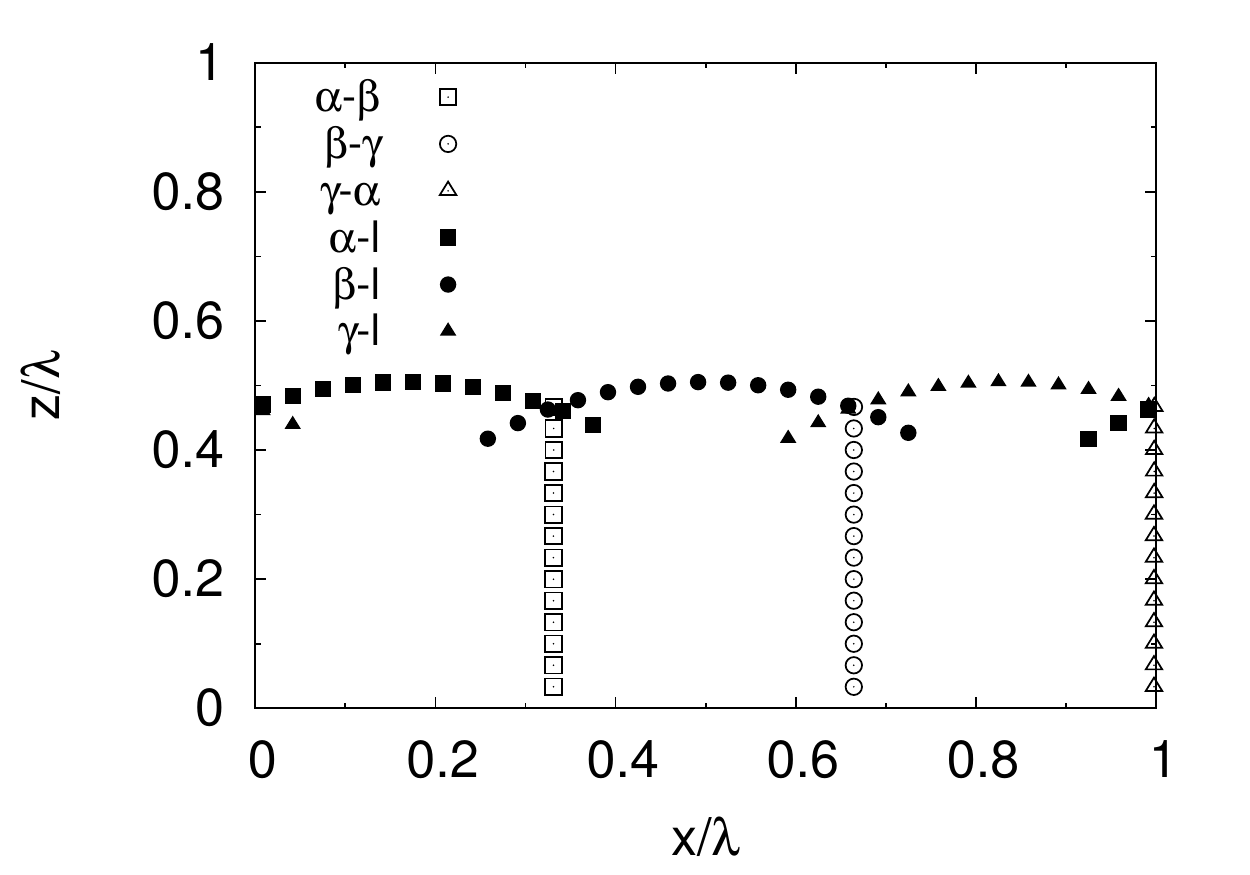}
\label{tilt_abg_2_2}
}\,
\subfloat[]{\includegraphics[width=0.7\linewidth]{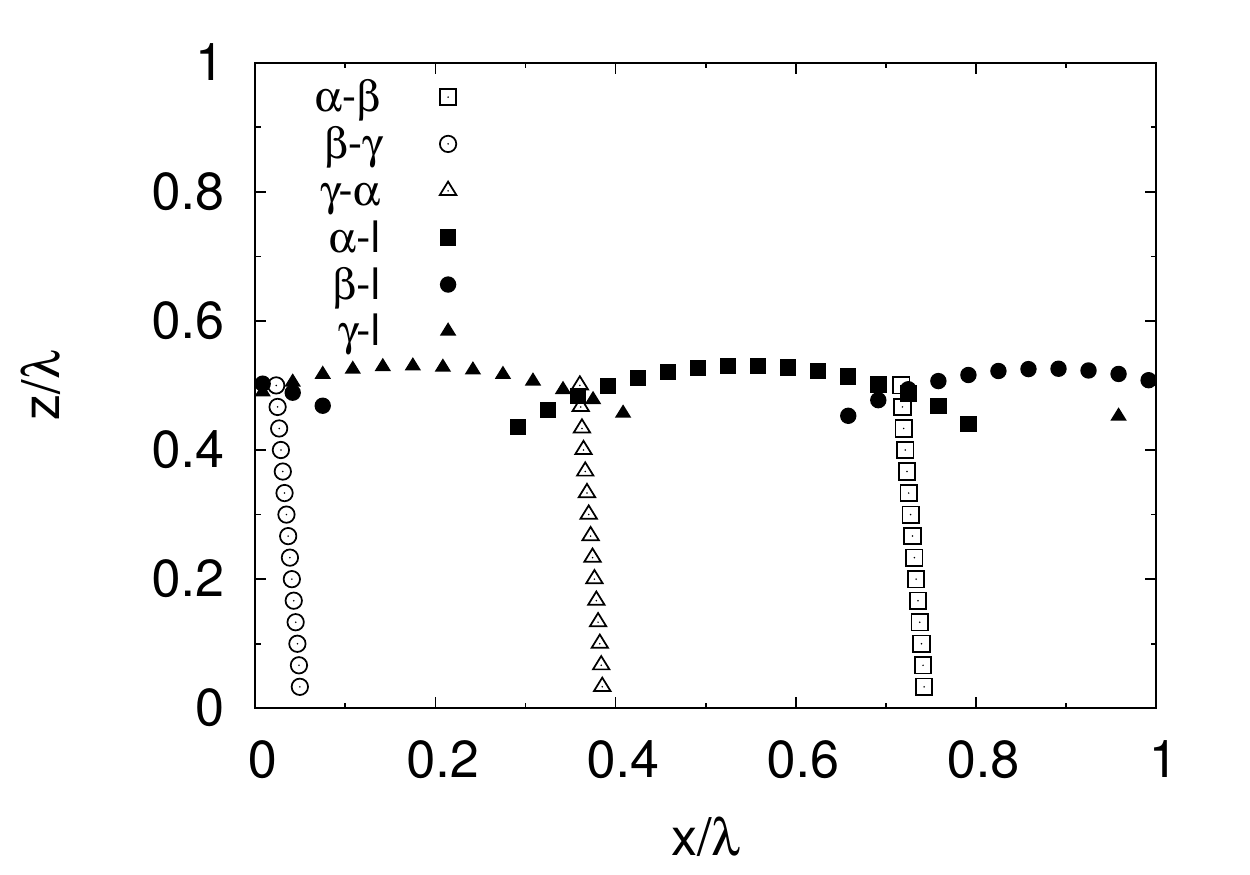}
\label{tilt_abg_2_1}
}
\caption{Plots showing orientations of all the interfaces for, (a)$D_{AA}=D_{BB}=2.0$, and 
(b)$D_{AA}=2.0, D_{BB}=1.0$, during three phase eutectic growth in a model symmetric ternary alloy. A single
 wavelength of the eutectic solids has the configuration $\alpha\beta\gamma$. }
\label{tilt_abg}
\end{figure}

However, for the case of unequal diffusivities, an 
inference from the phase-field simulations
can be seen in Fig.~\ref{tilt_abg} where we notice
a tilt in the lamellar arrangement with respect
to the growth direction. 

We note that this tilt is not an "instability" that occurs
beyond a spacing as has been reported during 
two and three-phase growth~\cite{karma1996,Bottin2016},
rather is a growth mode that is selected, which has also 
been found in previous three-phase growth simulations
although due to different conditions~\cite{Hecht2004}. Clearly, 
this is a prediction that is impossible to derive 
from the theoretical calculations that we present
in this paper, and is certainly a limitation of the 
applicability of such calculations. More elaborately, 
in order for the theoretical predictions to be effective, 
one must have the information about the steady-state
growth mode that is either derived experimentally, 
or through simulations. 

One of the reasons for the $\alpha\beta\gamma$
configurations to tilt is that there are no symmetry
planes once the volume fractions of the phases 
become unequal. The next arrangement $\alpha\beta\alpha\gamma$
however, possesses, two mirror axes, passing through 
the $\beta$ and the $\gamma$ phases. Going by the 
symmetry arguments placed in a previous paper
\cite{Choudhury2011}, a steady-state growth mode
where the lamellae are aligned with the growth 
direction, is therefore expected. We repeat
the simulation and analytical calculations for
this configuration for different diffusivities, 
and the results are reported in Figs.\ref{abag_contd}.
It is important to note that for this configuration
a short wavelength instability exists which results in the transformation 
of the $\alpha\beta\alpha\gamma$ to $\alpha\beta\gamma$ 
occurring below a critical wavelength (see discussion in \cite{Choudhury2011}).
Thereby, we limit our analysis to only the stable lamellar 
states. For these spacings, we again derive an excellent agreement
for the undercooling vs spacing variations, volume fractions
and the phase compositions, between phase-field
simulations and theoretical predictions. Due to the 
variation in the stability regimes we have limited our
calculations for the case of only unequal diffusivities 
$D_{AA}=1.0, D_{BB}=2.0$, as the stability region for
the contrary case of $D_{AA}=2.0, D_{BB}=1.0$ is very 
small.

\begin{figure}[!htbp]
\centering
\subfloat[]{\includegraphics[width=0.7\linewidth]{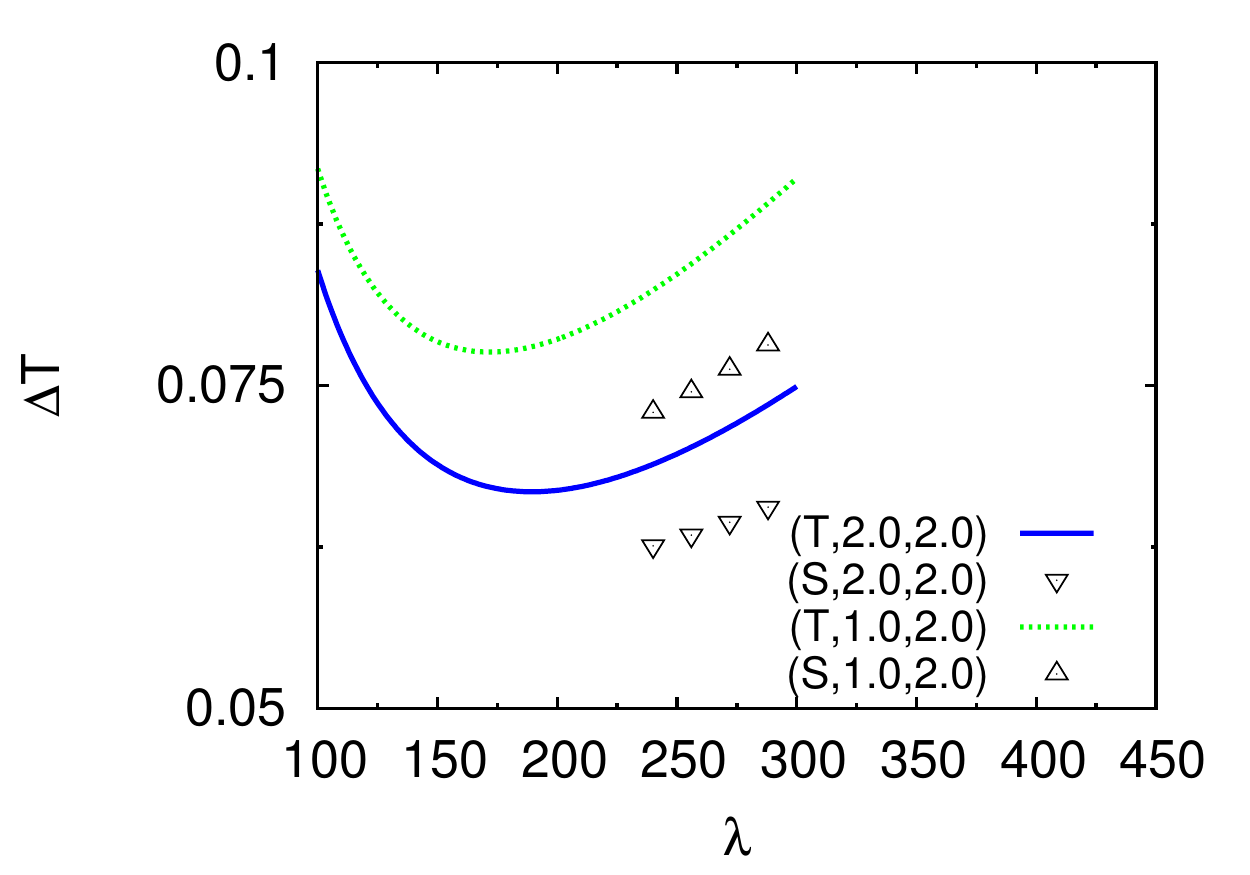}
\label{dtlam_abag}
}\,
\subfloat[]{\includegraphics[width=0.7\linewidth]{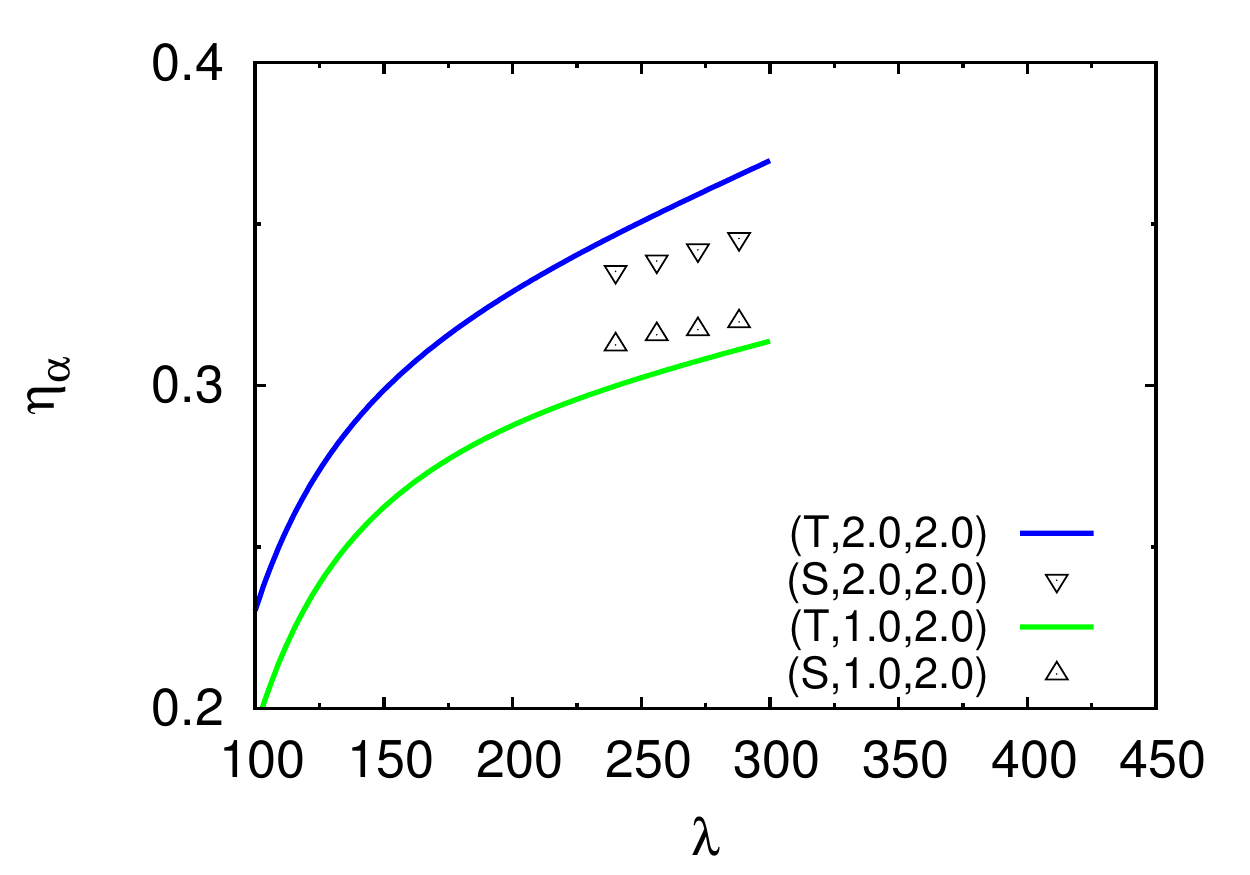}
\label{et_alp_lam_abag}
}\,
\subfloat[]{\includegraphics[width=0.7\linewidth]{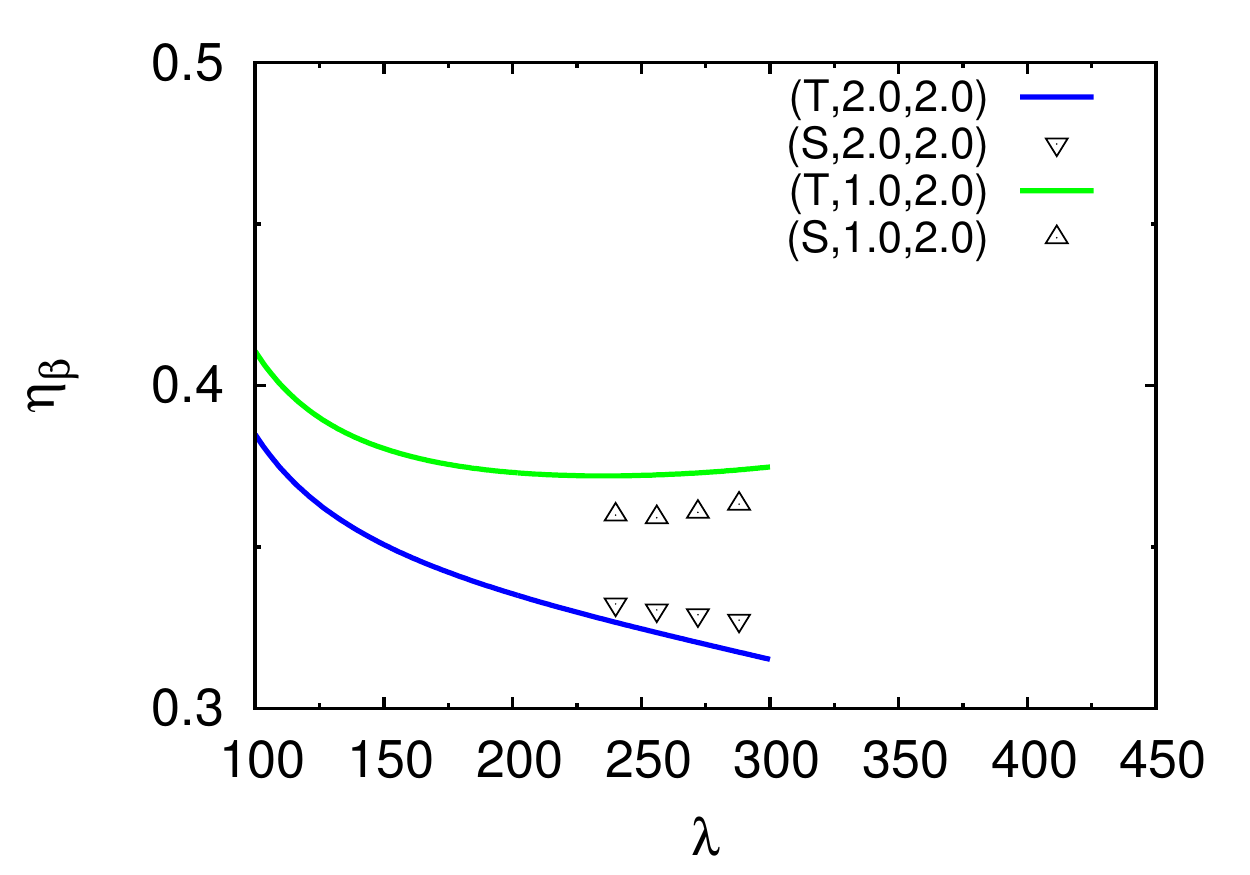}
\label{et_bet_lam_abag}
}
\caption{(Color online) Plots showing variations of (a)$\Delta T$, (b) $\eta_\alpha$, and (c) $\eta_\beta$,
with $\lambda$, during three phase eutectic growth in a model symmetric ternary alloy. A single
 wavelength of the eutectic solids has the configuration $\alpha\beta\alpha\gamma$. 
 The figure legends can be interpreted in the same way as described in the caption of
Fig.~\ref{symm_3030}.     
}
\label{abag}
\end{figure}

\begin{figure}[!htbp]
\centering
\ContinuedFloat
\subfloat[]{\includegraphics[width=0.7\linewidth]{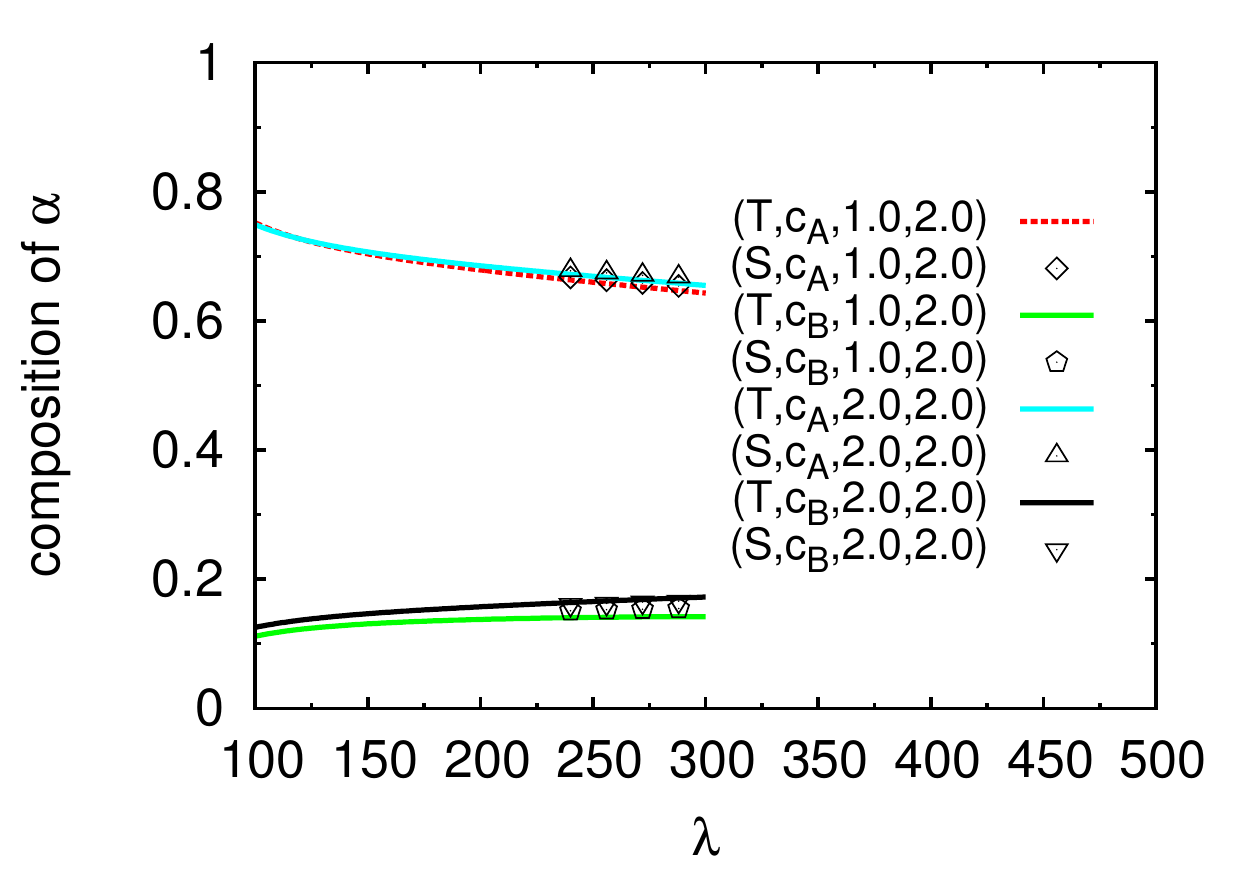}
\label{calp_abag}
}\,
\subfloat[]{\includegraphics[width=0.7\linewidth]{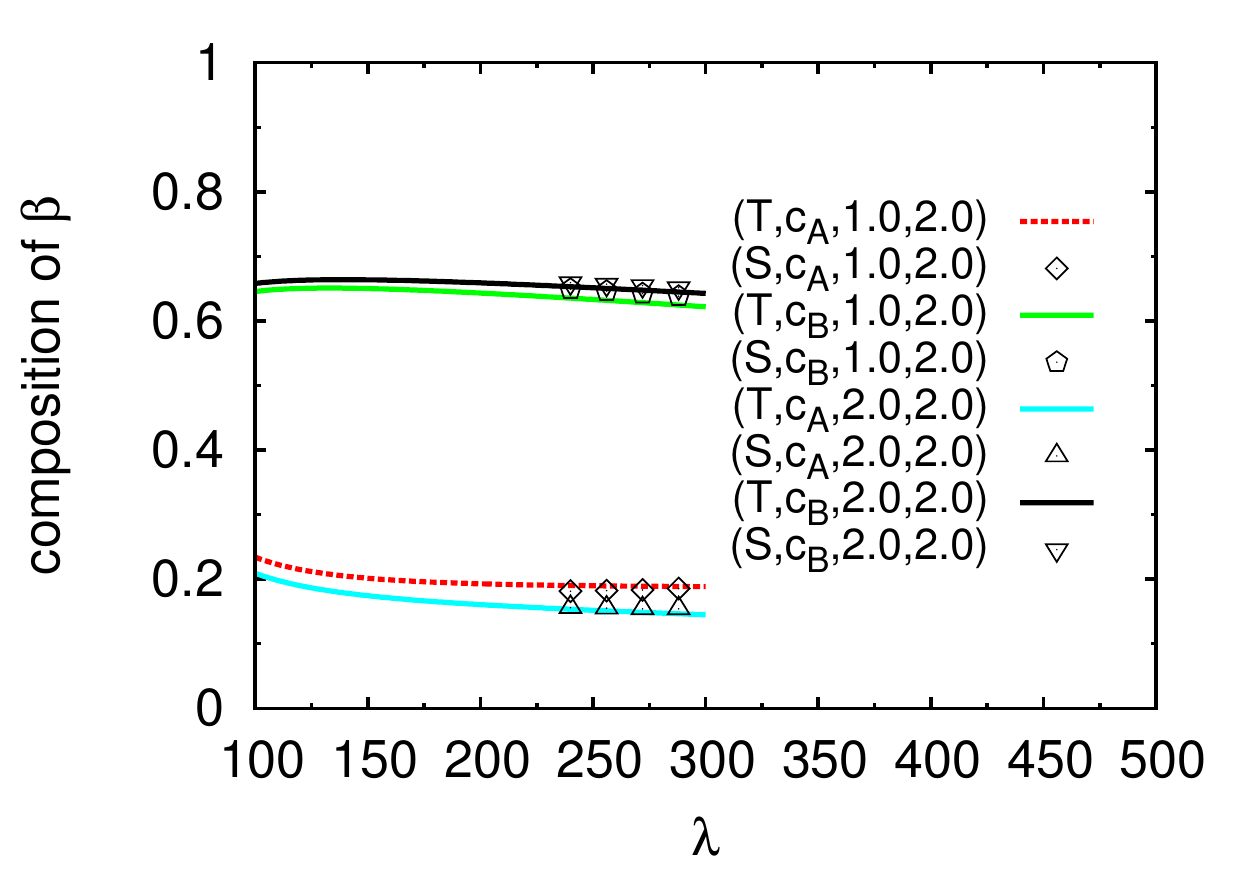}
\label{cbet_abag}
}\,
\subfloat[]{\includegraphics[width=0.7\linewidth]{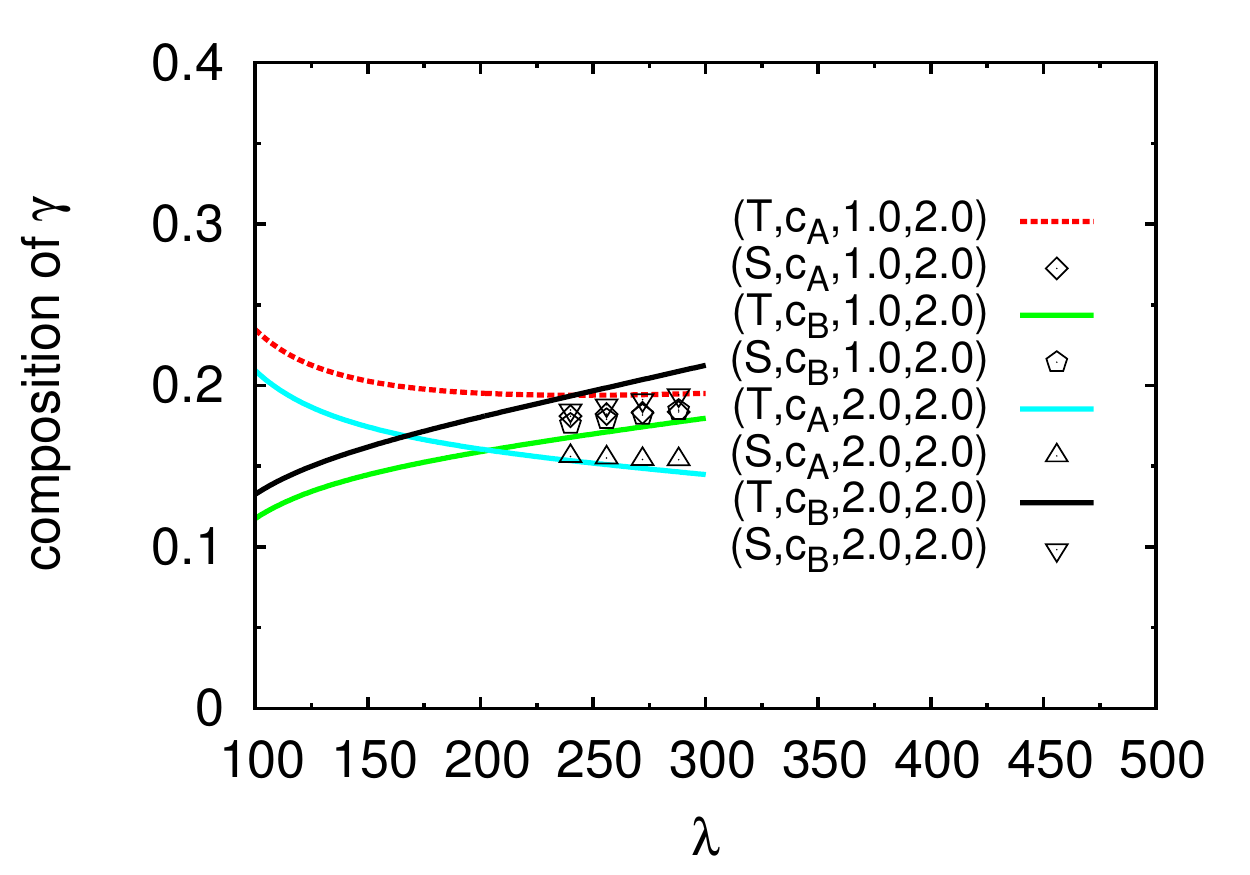}
\label{cgam_abag}
}
\caption{(Color online) Plots showing variations of (d) $\alpha$, (e)$\beta$, and (f)$\gamma$ phase compositions
with $\lambda$, in three phase eutectic growth in a model symmetric ternary alloy. A single
 wavelength of the eutectic solids has the configuration $\alpha\beta\alpha\gamma$. 
 The figure legends can be interpreted in the same way as described in the caption of
Fig.~\ref{symm_3030}.     
}
\label{abag_contd}
\end{figure}

\section{Conclusions}
In this study, we derive an analytical theory to determine the interfacial undercoolings, volume fractions 
and compositions of the solid phases in directionally solidifying lamellar eutectics for a generic multi-component,
multi-phase alloy. While our work bears similarities to the recently published work of Senninger
and Voorhees \cite{Senninger2016} which is particular for two-phase growth, our work gives
a generic prescription for treating any given multi-phase, multi-component alloy and in this
respect can be seen as an extension of the previous work in \cite{Choudhury2011}. A principal
point in our theoretical calculations is that we treat the multi-variant and invariant eutectic
reactions alike, by expressing the boundary layer compositions as functions of the respective 
state variables, which for our derivation are the diffusion potentials, the phase fractions 
and the undercooling. This allows us to solve the system of equations self-consistently for
the undercoolings, phase fractions and the phase compositions along with the boundary 
layer compositions irrespective of the degrees of freedom in the system. Our derivation,
thus unifies the method of theoretical calculations of the Jackson-Hunt type for any given multi-variant/invariant
eutectic growth.

We also perform phase-field simulations to corroborate our theoretical predictions and they are found to be in 
reasonably good match with each other where we investigate the case of monovariant two-phase and 
three-phase invariant growth. Both the phase-field and the analytical theory exhibit the same trends in the 
variation of interfacial undercooling, solid phase volume fractions and compositions with change in lamellar width.
It is important to highlight that the numerical differences in the predictions obtained from the two techniques are 
attributed to the assumption of a flat interface in the analytical calculations. Particularly, 
asymmetry in the interfacial shapes brought about either by strongly different phase fractions
or interfacial energies result in asymmetric discrepancies between the theoretical predictions
and the phase-field predictions. Thus we expect the match between the two methods w.r.t  
the predictions of the phase fractions and phase compositions to be the 
best for situations where the interfacial shapes of the phases are similar.
Furthermore, we note in passing that while the theoretical expressions are 
generic in the spirit in which they have been derived, the existence of a 
steady-state lamellar growth mode needs to be ascertained through either 
phase-field simulations or experiments, before applying the results.

Secondly our study clearly highlights the importance of understanding the 
dependence of phase fractions on the diffusivity matrices. 
Changes in volume fractions can be associated with microstructural changes during 
two-phase growth(lamellar to rod), and many further possibilities
during three-phase growth as seen in~\cite{Choudhury2016,Lahiri2015}.
Therefore, dependence of the volume fractions on the diffusivity 
matrices needs to be accounted for in order to derive a better understanding
of pattern formation during bulk eutectic growth in multi-phase systems.  

\section{Appendix}
The expression of the free energy of a solid phase($p$) is given by,
\begin{align}
f^p(\mathbf{c^p},T)=\sum_{i=1,j=1,i<=j}^{K-1,K-1} A_{ij}^p c_i^p c_j^p + \sum_{j=1}^{K-1} B_j^p(T) c_j^p + D^p(T),
\label{fsolid}
\end{align}
and that of the liquid is given by,
\begin{align}
 f^l(\mathbf{c^l})=\sum_{i=1,j=1,i<=j}^{K-1,K-1} A_{ij}^l c_i^l c_j^l,
 \label{fliquid}
\end{align}
where the constants $A_{ij}^{p,l}$ are set to obtain $\left[\dfrac{\partial c^{p,l}}{\partial \mu}\right]$ matrices 
while $B_j^p$ and $D^p$ are determined from the equilibrium between solid and the liquid phases at a particular temperature
as described in~\cite{Choudhury2015}.

\bibliography{gen_jh}
\end{document}